\documentclass[aip,jmp,amsmath,amssymb,reprint]{revtex4-1}

\usepackage{enumitem}
\usepackage{array}
\usepackage{float}
\usepackage{graphicx}
\usepackage{dcolumn}
\usepackage{bm}

\makeatletter

\newcommand{\Rmnum}[1]{\expandafter\@slowromancap\romannumeral #1@}
\makeatother

\begin{document}

\preprint{AIP/123-QED}

\title{Wave modelling in a cylindrical non-uniform helicon discharge}

\author{L. Chang}
\email{chang.lei@anu.edu.au.}
\author{M. J. Hole}
\author{J. F. Caneses}
\affiliation{Plasma Research Laboratory, Research School of Physics and Engineering, Australian National University, Canberra, ACT 0200, Australia}

\author{G. Chen}
\affiliation{Oak Ridge National Laboratory, Oak Ridge, Tennessee 37831, USA}

\author{B. D. Blackwell}
\author{C. S. Corr}
\affiliation{Plasma Research Laboratory, Research School of Physics and Engineering, Australian National University, Canberra, ACT 0200, Australia}

\begin{abstract}
A radio frequency (RF) field solver based on Maxwell's equations and a cold plasma dielectric tensor is employed to describe wave phenomena observed in a cylindrical non-uniform helicon discharge. The experiment is carried out on a recently built linear plasma-material interaction machine: the MAGnetized Plasma Interaction Experiment (MAGPIE) [B. D. Blackwell, J. F. Caneses, C. Samuell, J. Wach, J. Howard, and C. S. Corr, submitted on 25 March 2012 to Plasma Sources Science and Technology], in which both plasma density and static magnetic field are functions of axial position. The field strength increases by a factor of $15$ from source to target plate, and plasma density and electron temperature are radially non-uniform. With an enhancement factor of $9.5$ to the electron-ion Coulomb collision frequency, $12\%$ reduction in the antenna radius, and the same other conditions as employed in the experiment, the solver produces axial and radial profiles of wave amplitude and phase that are consistent with measurements. Ion-acoustic turbulence, which can happen if electron drift velocity exceeds the speed of sound in magnetized plasmas, may account for the factor of $9.5$ used to match simulated results with experimental data. To overcome the single $m$ vacuum solution limitations of the RF solver, which can only compute the glass response to the same mode number of the antenna, we have adjusted the antenna radius to match the wave field strength in the plasma. A numerical study on the effects of axial gradient in plasma density and static magnetic field on wave propagations is performed, revealing that the helicon wave has weaker attenuation away from the antenna in a focused field compared to a uniform field. This may be consistent with observations of increased ionization efficiency and plasma production in a non-uniform field. We find that the relationship between plasma density, static magnetic field strength and axial wavelength agrees well with a simple theory developed previously. A numerical scan of the enhancement factor to the electron-ion Coulomb collision frequency from $1$ to $15$ shows that the wave amplitude is lowered and the power deposited into the core plasma decreases as the enhancement factor increases, possibly due to the stronger edge heating for higher collision frequencies.
\end{abstract}

\maketitle

\section{introduction}
Generically, a helicon discharge usually refers to a cylindrical plasma discharge with an axial static magnetic field, driven by radio frequency (RF) waves at frequencies between the ion and electron cyclotron frequencies, $\omega_{ci} \ll \omega \ll \omega_{ce}$.\cite{Carter:2002aa} A helicon discharge produces plasmas with densities typically much higher than capacitive and inductive plasma sources operating at similar pressures and input RF powers.\cite{Scime:2008aa} Because of this high ionization efficiency, helicon discharges have found applications in various fields, including: plasma rocket propulsion,\cite{Arefiev:2004ab, Ziemba:2005aa, Charles:2009aa, Batishchev:2009aa} a plasma source for magnetic fusion studies,\cite{Loewenhardt:1991aa} Alfv\'{e}n wave propagation,\cite{Hanna:2001aa} RF current drive,\cite{Petrzilka:1994aa} laser plasma sources,\cite{Zhu:1989aa} possibly semiconductor processing, electrodeless beam sources, and laser accelerators.\cite{Chen:1996ab}

To date, most helicon studies have treated devices with uniform static magnetic fields, however, many applications require operation with axial magnetic field variations. \cite{Mori:2004aa} A few researchers have investigated helicon plasma sources with non-uniform magnetic fields, and have found that the plasma density increased when a cusp or non-uniform magnetic field was placed in the vicinity of the helicon antenna.\cite{Boswell:1997aa, Chen:1992aa, Chen:1997aa, Gilland:1998aa} However, detailed examination of the reasons for this enhanced plasma density has not yet been conducted, although fast electrons and improved confinement are mentioned as possible contributors. Guo et al.\cite{Guo:1999aa} furthered this study by looking at the effects of non-uniform magnetic field on source operations, and found that strong axial gradient in density associated with non-uniform field configuration can contribute to the absorption of wave fields and a high ionization efficiency. Takechi et al. \cite{Takechi:1999aa} also suggested that there may be a close relationship between plasma density profile and RF wave propagation and absorption regions, finding the density uniformity in the radial direction improved markedly with the cusp field. Therefore, studying the effects of various static magnetic field configurations on helicon wave propagation is of significant importance to producing desired plasma profiles and understanding the role of magnetic field in helicon plasma generations. 

This paper is dedicated to modelling the wave field observed in MAGPIE (MAGnetized Plasma Interaction Experiment), and investigating helicon wave propagation in the non-uniform magnetized plasma of this machine, in which both the static magnetic field and its associated plasma density are functions of axial position. The plasma density and electron temperature are also dependent on radius. We assume in this study that the electron temperature is independent of $z$ and the static magnetic field is almost independent of $r$ (Eq. (7)). MAGPIE is a linear plasma-material interaction machine which was recently built in the Plasma Research Laboratory at the Australian National University, and designed for studying basic plasma phenomena, testing materials in near-fusion conditions, and developing potential diagnostics applicable for the edge regions of a fusion reactor.\cite{Blackwell:2012aa} A RF field solver, \cite{Chen:2006aa} based on Maxwell's equations and a cold plasma dielectric tensor, is employed in this study. The motivations of our work are to explain the wave field measurements in MAGPIE, and to study the effects of magnetic field configuration on helicon wave propagation. The rest of the paper is organized as follows: Section \Rmnum{2} describes the experimental apparatus and diagnostic tools, together with the measured static magnetic field, plasma density and temperature profiles; Section \Rmnum{3} provides an overview of the employed theoretical model and the numerical code, together with comparisons between computed and measured wave fields; Section \Rmnum{4} is dedicated to a numerical study of the effects of plasma density and static magnetic field profiles on the wave propagation characteristics; Section \Rmnum{5} aims to study the physics meaning of the enhancement factor to electron-ion Coulomb collision frequency, and the effects of the direction of static magnetic field on wave propagations;\cite{Chen:2006aa, Zhang:2008aa, Lee:2011aa} finally, Section \Rmnum{6} presents concluding remarks and future work for continuing research. 

\section{experiment}
\subsection{Experimental setup}
\label{subsec: antenna}
\begin{figure*}
\begin{center}$
\begin{array}{c}
\includegraphics[width=0.9\textwidth,angle=0]{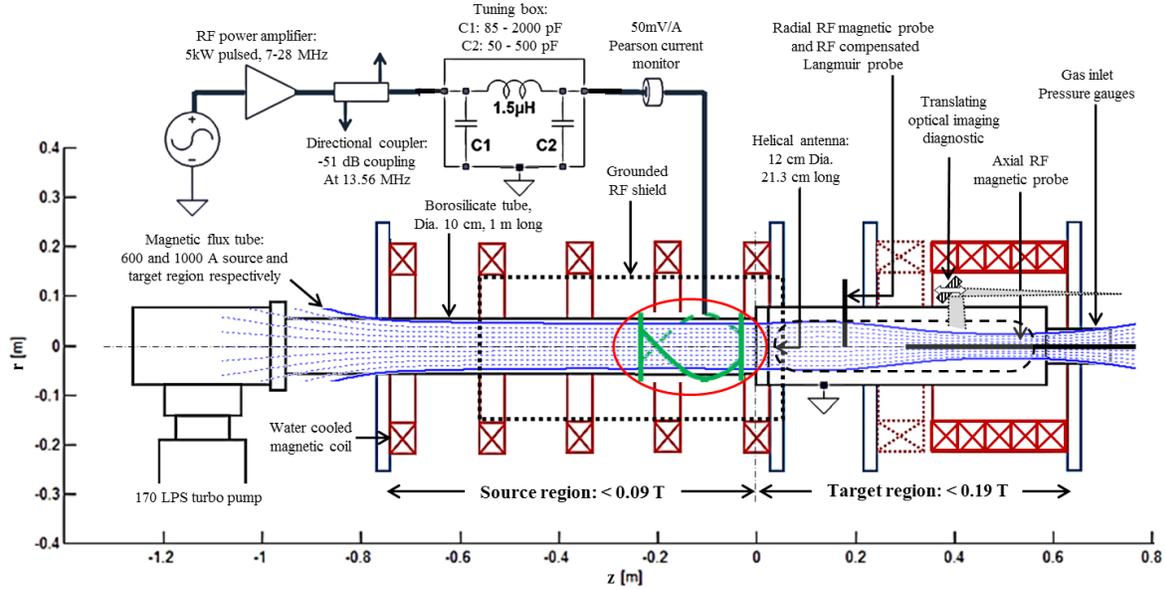}
\end{array}$
\end{center}
\caption{A schematic of the MAGPIE (MAGnetized Plasma Interection Experiment).\cite{Blackwell:2012aa} A circle denotes the position of the helicon antenna which is left hand half-turn helical. The dot-dashed line is the machine and coordinate system axis, defining $r=0$~m. The coordinate system is right-handed with $\theta=0$ chosen to be the zenith angle. }
\label{mdf}
\end{figure*}

Similar to other helicon devices,\cite{Carter:2002aa} MAGPIE mainly consists of a dielectric glass tube surrounded by an antenna, a vacuum pumping system, and a gas feeding system, together with a power supply system connected to the antenna, and various diagnostics. Figure 1 shows a schematic and introduces a cylindrical (r, $\theta$, z) coordinate system.\cite{Blackwell:2012aa} The plasma is formed in the region under the antenna ($-0.243<z<-0.03$~m) and the near field to the antenna.\cite{Chen:1996aa} Following convention, however, we define the whole glass tube ($-1<z<0$~m) as the source region and the compressed field region ($0<z<0.7$~m) as the target region (or equivalently ``diffusion region" in some references). In MAGPIE, the $z<-0.243$~m region is named ``upstream" and $z>-0.03$~m ``downstream".  

A glass tube of length $1$~m and radius $0.05$~m is used to contain source plasmas in MAGPIE. A left hand half-turn helical antenna, $0.213$~m in length and $0.06$~m in radius, is wrapped around the tube and connected to a tuning box which can be adjusted between $7$ and $28$~MHz, a directional coupler, a $5$~kW RF amplifier,  and a $150$~W pre-amplifying unit. For the present study, an RF power of $2.1$~kW, $13.56$~MHz, pulse width of $1.5$~ms and duty circle of $1.5$~\% is used. The antenna current is measured by a Rogowski-coil-type current monitor. For these experiments an antenna current of magnitude $I_\text{a}=38.8$~A was measured. A grounded stainless steel cylindrical mesh surrounding the whole source region is employed to protect users. The source region is connected on-axis to the aluminium target chamber which is $0.7$~m in length and $0.08$~m in radius. Gases are fed through the downstream end of the target chamber, and drawn to the upstream end of the source tube by a $170$~L/s turbo pump. Gas pressures are measured in the target chamber by a hot cathode Bayard-Alpert Ionization gauge ($<0.01$~Pa), a Baratron pressure gauge ($0.01$--$10$~Pa) and a Convectron ($0.1$~Pa--$101.33$~kPa) for pressure process. In this experiment, argon gas is used with a filling pressure of $P_\text{B}=0.41$~Pa. The two regions, source and target, are surrounded by a set of water cooled solenoids, with internal radius of $0.15$~m. These source and target sets of solenoids are powered by two independent $1000$~A, $20$~V DC power supplies, providing flexibility in the axial configuration of the static magnetic field, e.~g. maximum of $0.09$~T and $0.19$~T in the source and target regions, respectively. The non-uniform field configuration is expected to provide a flexible degree of radial confinement, better plasma transport from the source tube to the target chamber, and possible increased plasma density according to previous studies.\cite{Blackwell:2012aa, Boswell:1997aa, Chen:1992aa, Chen:1997aa, Gilland:1998aa, Guo:1999aa} The direction of the static magnetic field employed in the present work points from target to source. 

\subsection{Plasma profile diagnostics}

\begin{figure}
\begin{center}$
\begin{array}{l}
(a)\\
\includegraphics[width=0.45\textwidth]{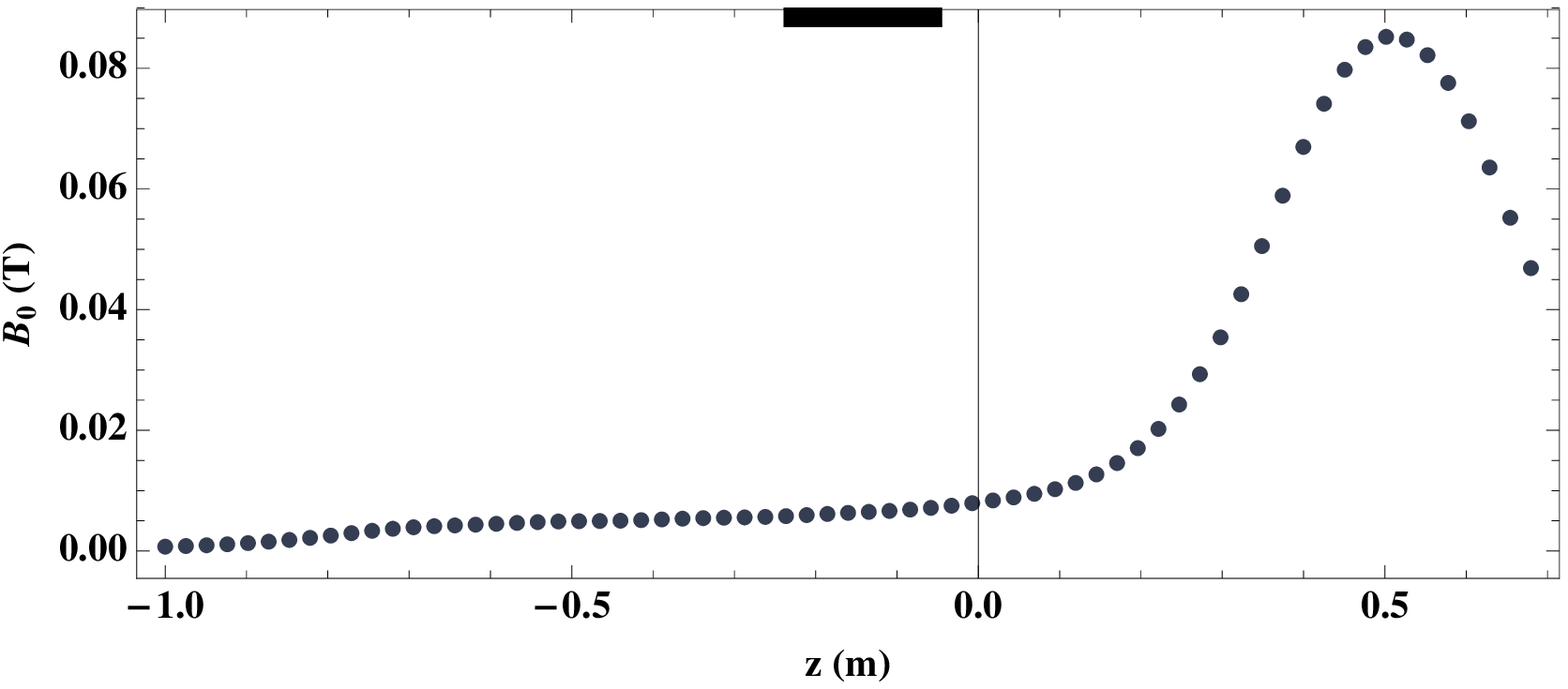}\\
(b)\\
\hspace{-0.02cm}\includegraphics[width=0.482\textwidth]{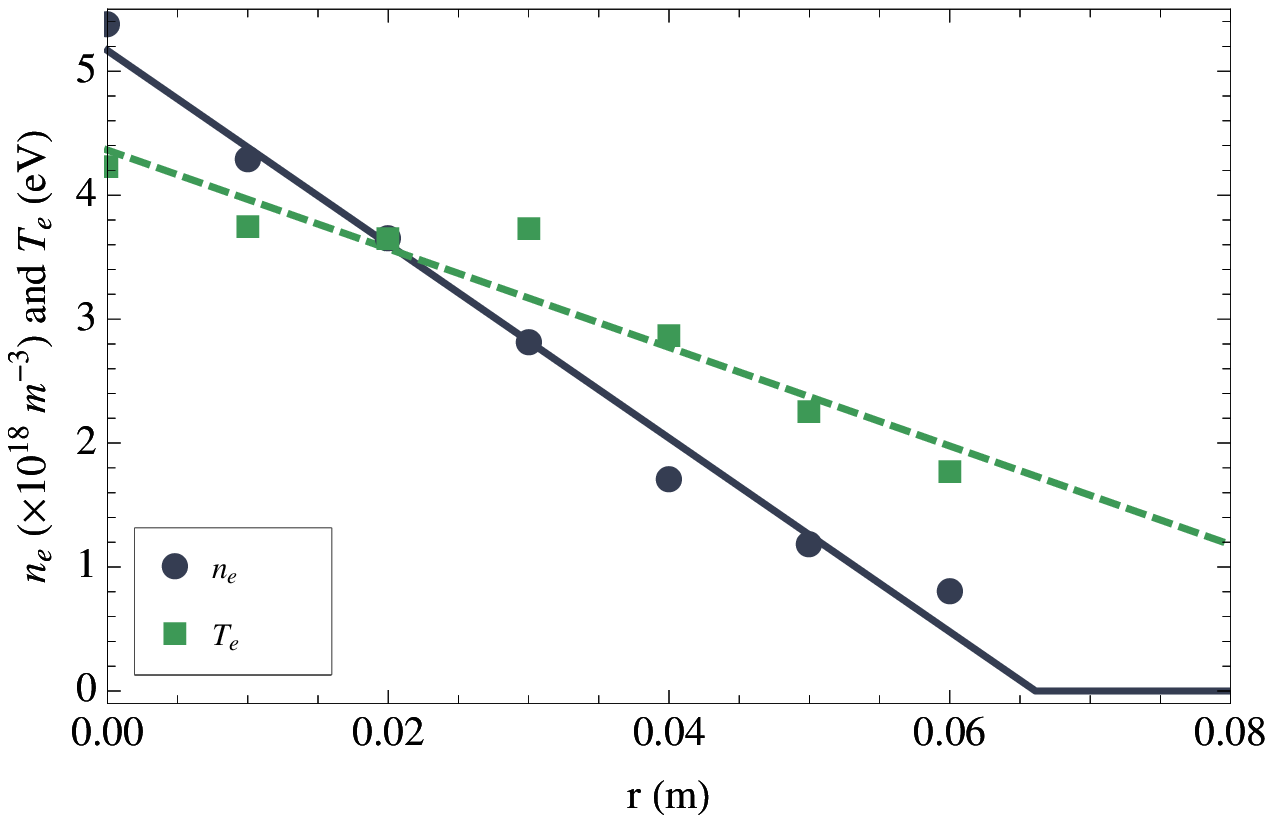}
\end{array}$
\end{center}
\caption{Typical measured profiles: (a) axial profile of static magnetic field on axis, (b) radial profiles of plasma density (dots) and electron temperature (squares) at $z=0.17$~m, together with their fitted lines, solid and dashed, respectively. The solid bar in (a) denotes the antenna location. }
\label{profiles}
\end{figure}

A passively compensated Langmuir probe was employed in our experiment to measure the plasma density and electron temperature, calculated from the $I(V)$ curve obtained by an Impedans Data Acquisition system.\cite{:aa} The probe comprises a platinum wire of diameter $0.1$~mm, and a surrounding alumina insulator. The length of the insulator is $6$~mm shorter than that of the platinum wire so that the exposed platinum wire forms the probe tip. Electron currents were drawn to clean the probe during regular intervals of argon discharges. The probe is located at $z=0.17$~m as shown in Fig. 1.

Typical measured axial profile of field strength and radial profiles of plasma density and electron temperature in MAGPIE are shown in Fig. 2. Particularly, as shown in Fig. 2(a), the increase in field strength ($B_0(z)$) from antenna end ($z=-0.243$~m) to field peak ($z=0.51$~m) is a factor of $15$. The axial profile of plasma density ($n_e(z)$) is assumed to be proportional to $B_0(z)$, consistent with generally accepted knowledge that the density follows the magnetic field linearly.\cite{Lieberman:2005aa, Boswell:aa} Figure 2(b) shows the radial profiles of plasma density ($n_e (r)$) and electron temperature ($T_e(r)$), measured at $z=0.17$~m and fitted with straight lines. During the density fitting procedure, in order to avoid negative fitted values, the density was set to zero in the region of $0.066\leq r \leq 0.08$~m. We assume the total density profile is separable, $n_e(r, z)=n_e(r)\times n_e(z)$. The fitted lines in $n_e(r)$ and $T_e(r)$, and the measured $B_0(z)$ data will be used in section \Rmnum{3} to constrain wave field simulations. 

\subsection{Wave field diagnostics}
Helicon wave fields were measured by a 2-axis ``B dot" or Mirnov probe. Details about the probe can be found in Blackwell et al..\cite{Blackwell:2012aa} To measure the axial profiles of $B_r$ and $B_z$, the probe was inserted on axis from the end of the target chamber. The probe is long enough to measure $B_r$ and $B_z$ in the range $-0.25<z<0.7$~m. Two perpendicular magnetic field components ($B_r$ and $B_z$ in this case) can be sampled simultaneously. To measure the radial profiles of the three magnetic wave components, $B_r$, $B_\theta$ and $B_z$, the probe was inserted radially at $z=0.17$~m, and rotated about its axis to measure $B_\theta$ and $B_z$. The B-dot probe couples inductively to the magnetic components of the helicon wave and electrostatically to the RF time varying plasma potential. To limit our measurements to the inductively coupled response, a current balun was employed to screen the electrostatic response. Further information about the procedure to eliminate the electrostatic response of the probe can be found in Franck et al..\cite{Franck:2002aa} Both axial and radial profiles of wave phase were measured through a phase-comparison method, similar to Light et al..\cite{Light:1995aa} To measure the variation in wave phase with axial position, the signal from an on-axis axially inserted probe was compared to the phase of the antenna current. A similar procedure was conducted to measure the variation in wave phase with radial position at $z=0.17$~m. It should be noted that all probe diagnostics are intrusive, and can affect the plasma parameters, and hence the wave fields. 

\section{simulation}
An RF field solver (or ElectroMagnetic Solver, EMS)\cite{Chen:2006aa} based on Maxwell's equations and a cold plasma dielectric tensor is employed in this study to interpret the RF waves measured in MAGPIE. This solver has been used successfully in explaining wave phenomena in two other machines: a helicon discharge machine at The University of Texas at Austin \cite{Lee:2011aa} and the LArge Plasma Device (LAPD) at the University of California at Los Angles.\cite{Zhang:2008aa} Details of the solver can be found in Chen et al.,\cite{Chen:2006aa} while a brief overview is given below. 

\subsection{Theoretical model}
The Maxwell's equations that this solver employs to determine the RF wave field in a helicon discharge are written in the frequency domain

\begin{equation}
\bigtriangledown\times\mathbf{E}=i \omega\mathbf{B}, 
\end{equation}

\begin{equation}
\frac{1}{\mu_0}\bigtriangledown\times\mathbf{B}=-i \omega \mathbf{D}+\mathbf{j_a}, 
\end{equation}
where $\mathbf{E}$ and $\mathbf{B}$ are the electric and magnetic fields, respectively, $\mathbf{D}$ is the electric displacement vector, $\omega$ is the antenna driving frequency, and $\mathbf{j_a}$ is the antenna current density. The quantities $\mathbf{D}$ and $\mathbf{E}$ are linked to each other by a dielectric tensor $\varepsilon_{\alpha\beta}$ that represents vacuum, glass and plasma. In the vacuum and glass regions, the dielectric tensor is $\varepsilon_{\alpha\beta}\equiv\varepsilon_\ast(r, z) \delta_{\alpha\beta}$, where $\delta_{\alpha\beta}$ is the Kronecker symbol and $\varepsilon_\ast(r, z)$ is a scalar. The term $\varepsilon_\ast(r, z)$ equals to $1$ and $\varepsilon_g$ for vacuum and glass regions, respectively, where $\varepsilon_g$ is the dielectric constant of glass. In the plasma region, because of the cold plasma approximation made here, the relation between $\mathbf{D}$ and $\mathbf{E}$ is in form of\cite{Stix:1992aa, Vincena:1999aa}

\begin{equation}
\mathbf{D}=\varepsilon_0(\varepsilon\mathbf{E}+ig[\mathbf{E}\times\mathbf{b}]+(\eta-\varepsilon)(\mathbf{E}\cdot\mathbf{b})\mathbf{b}),
\end{equation}
where $\mathbf{b}\equiv\mathbf{B_0}/B_0$ is the unit vector along the static magnetic field and 

\begin{equation}
\varepsilon=1-\sum\limits_{\alpha}\frac{\omega+i\nu_\alpha}{\omega}\frac{\omega^2_{p\alpha}}{(\omega+i\nu_\alpha)^2-\omega^2_{c\alpha}}, 
\end{equation}

\begin{equation}
g=-\sum\limits_{\alpha}\frac{\omega_{c\alpha}}{\omega}\frac{\omega^2_{p\alpha}}{(\omega+i\nu_\alpha)^2-\omega^2_{c\alpha}}, 
\end{equation}

\begin{equation}
\eta=1-\sum\limits_{\alpha}\frac{\omega^2_{p\alpha}}{\omega(\omega+i\nu_\alpha)}. 
\end{equation}
Here the subscript $\alpha$ labels particle species, i. e. electron and ion, $\omega_{p\alpha}\equiv\sqrt{n_\alpha q_\alpha^2/\varepsilon_0 m_\alpha}$ is the plasma frequency, $\omega_{c\alpha}\equiv q_\alpha B_0/m_\alpha$ gyrofrequency, and $\nu_\alpha$ collision frequency between species. The plasma is assumed to be nearly fully ionized in the present study, so that neutral collisions are neglected. Because $\nu_{ee}$ and $\nu_{ii}$ do not contribute to the momentum exchange between electron and ion fluids,\cite{Chen:1984aa} collision frequencies for electrons and ions species are $\nu_{e}=\nu_{ei}=2.91\times 10^{-12}n_e T^{-3/2}_e \text{ln}\Lambda$ and $\nu_{i}=\nu_{ie}=m_e m_i^{-1} \nu_{ei}$, respectively, from which we can see $\nu_{ie}\ll\nu_{ei}$. Here, $T_e$ and $n_e$ are given in eV and $\rm{m^{-3}}$, respectively, and the Coulomb logarithm is calculated to be ln$\Lambda=12$. Singly ionized argon ions are assumed in this study, so that $q_i=-q_e=|e|$.

The externally applied $B_0(r, \theta, z)$ is assumed to be axisymmetric, with $B_{0r}\ll B_{0z}$ and $B_{0\theta}=0$. Therefore, it is appropriate to use a near axis expansion\cite{Zhang:2008aa} for $B_0(r, \theta, z)$, namely $B_{0z}$ is only dependent on $z$ and 

\begin{equation}
B_{0r}(r,~z)=-\frac{1}{2}r\frac{\partial B_{0z} (z)}{\partial z}. 
\end{equation}

The antenna, as described in section~\ref{subsec: antenna}, is a left hand half-turn helical antenna. We assume that the antenna current is divergence free, to eliminate the capacitive coupling. Fourier components of the antenna current density are given by

\begin{equation}
j_{ar}=0, 
\end{equation}

\begin{equation}
\setlength{\extrarowheight}{0.3cm}
\begin{array}{cc}
j_{a\theta}=&I_a\frac{e^{i m \pi}-1}{2}\delta(r-R_a)(\frac{i}{m \pi}[\delta(z-z_a)+\delta(z-z_a-L_a)] \\
&+\frac{H(z-z_a)H(z_a+L_a-z)}{L_a}e^{-i m \pi[1+(z-z_a)/L_a]}), 
\end{array}
\end{equation}

\begin{equation}
\setlength{\extrarowheight}{0.3cm}
\begin{array}{cc}
j_{az}=&I_a\frac{e^{-i m \pi[1+(z-z_a)/L_a]}}{\pi R_a}\frac{1-e^{i m \pi}}{2}\delta(r-R_a)\\
&\times H(z-z_a)H(z_a+L_a-z), 
\end{array}
\end{equation}
where $L_a$ is the antenna length, $R_a$ the antenna radius, $z_a$ the distance between the antenna and the endplate in the source region, and $H$ the Heaviside step function. Note that the antenna geometry selects only odd harmonic mode number $m$, as indicated by Chen et al..\cite{Chen:2006aa} 

\subsection{Boundary conditions}

For a given azimuthal mode number $m$, Eq. (1) and Eq. (2) are first Fourier transformed with respect to the azimuthal angle, and then solved through a finite difference scheme on a 2D domain $(z,~r)$, as shown in Fig. 3. In the experiment, there is a radial air gap ( $0.055<r<0.0585$~m) between the antenna and the glass tube, which is taken as glass region in the computational domain. We found that simulated results are insensitive to the dielectric constant in the glass region $0.05<r<0.055$~m by varying the constant from $1$ to $10$ and no change detected in the wave field, therefore, we expended the glass area radially to fill this air gap. The thickness of the antenna is approximately $0.002-0.003$~m. 

\begin{figure}[h]
\begin{center}$
\begin{array}{c}
\includegraphics[width=0.475\textwidth,angle=0]{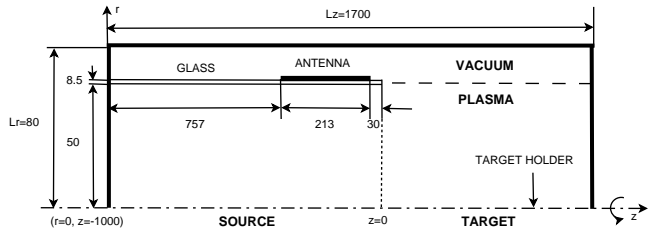}
\end{array}$
\end{center}
\caption{Computational domain employed to simulate the experimental setup shown in Fig. 1. Here all dimensions are given in millimetres. }
\label{domain}
\end{figure}

The radial wall of the target chamber and the axial endplates are ideally
conducting so that the tangential components of $\mathbf{E}$ vanish at the surface of these boundaries, i.e., 

\begin{equation}
E_\theta(L_r, z)=E_z(L_r, z)=0, 
\end{equation}

\begin{equation}
E_r(r, 0)=E_\theta(r, 0)=0, 
\end{equation}

\begin{equation}
E_r(r, L_z)=E_\theta(r, L_z)=0,  
\end{equation}
where $L_r$ and $L_z$ are the radius of the target chamber and the length of the whole machine, respectively. Moreover, all field components must be regular on axis, thus, $B_\theta |_{r=0}=0$ and $(rE_\theta)|_{r=0}=0$ for $m=0$; $E_z|_{r=0}$ and $(rE_\theta)|_{r=0}$ for $m\neq 0$.\cite{Zhang:2008aa} In the present work, we choose the fundamental odd mode number $m=1$, which is preferentially excited in the helicon discharge launched by a left hand half-turn helical antenna.\cite{Chen:1996aa, Light:1995aa, Light:1995ab}

\subsection{Computed and measured wave fields}
\label{subsec: standing}

\begin{figure}
\begin{center}$
\begin{array}{l}
(a)\\
\hspace{0.17cm}\includegraphics[width=0.522\textwidth,angle=0]{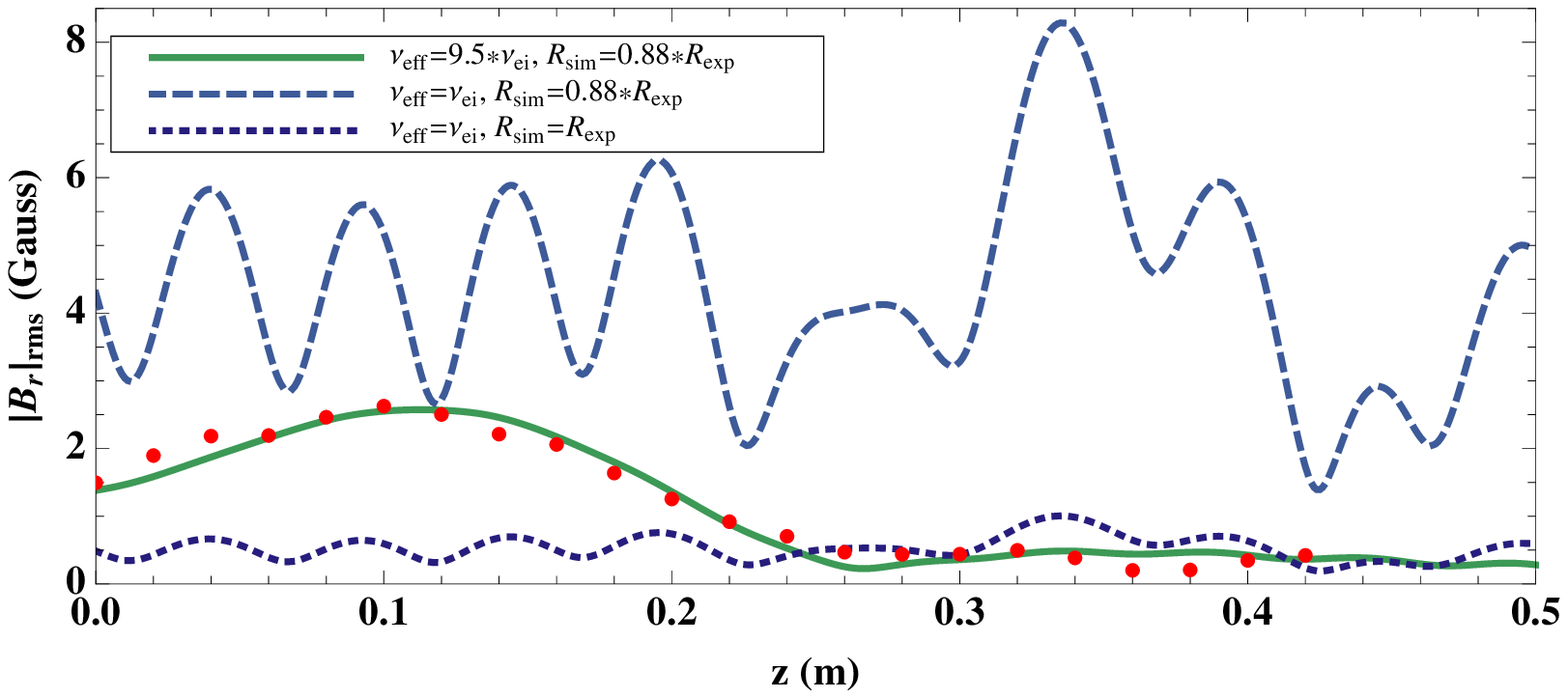}\\
(b)\\
\hspace{-0.14cm}\includegraphics[width=0.48\textwidth,angle=0]{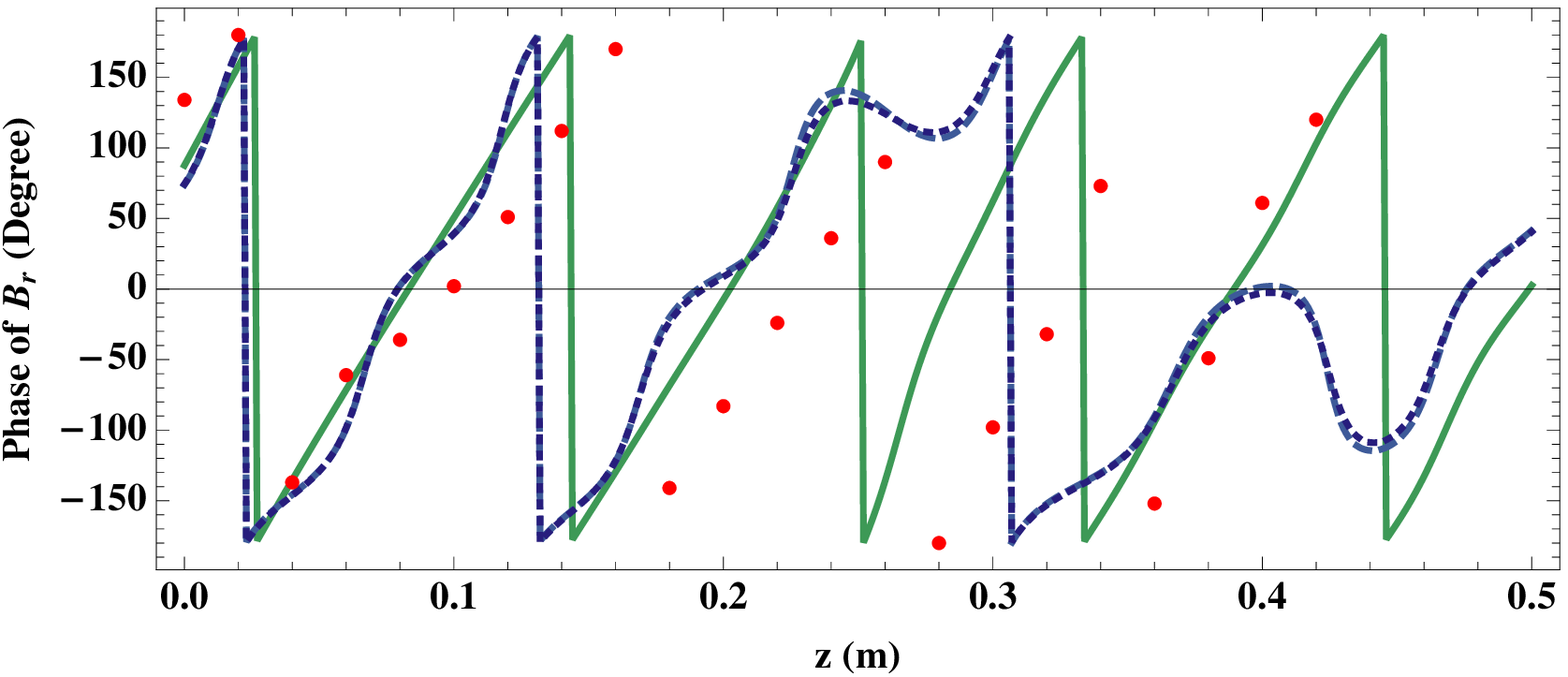}
\end{array}$
\end{center}
\caption{Variations of magnetic wave field in axial direction (on-axis): (a) $|B_r|_{\text{rms}}$, (b) phase of $B_r$. Computed results (lines: dotted for $\nu_{\text{eff}}=\nu_{ei}$ and $R_{\text{sim}}=R_{\text{exp}}$, dashed for $\nu_{\text{eff}}=\nu_{ei}$ and $R_{\text{sim}}=0.88R_{\text{exp}}$, and solid for $\nu_{\text{eff}}=9.5\nu_{ei}$ and $R_{\text{sim}}=0.88R_{\text{exp}}$) are compared with experimental data (dots). }
\label{axial}
\end{figure}

Based on the measured field strength configuration and plasma profiles shown in Fig. 2, simulations are performed. Figure 4 shows the axial profiles of the computed $B_r$ amplitude and phase on axis, and their comparisons with experimental data. With the collisionality set to $\nu_{\text{eff}}=\nu_{ei}$, where $\nu_{\text{eff}}$ is the effective collision frequency, and the antenna radius set to match the experiment, the predicted wave field is $\sim 30\%$ of the measured value, and the profile a poor match to the experiment. It is possible to obtain better agreement by varying the collisionality, which strongly affects the profile but leaves the magnitude largely unchanged, and the antenna radius, which strongly affects the field amplitude and leaves the radial and axial profiles of $\mathbf{B}$ unchanged. A qualitative match between measurement and simulation of the axial variation of $B_r$ is found using an enhancement in collisionality of $\nu_{\text{eff}}=\zeta (\nu_{ei}+\nu_{ie})\approx \zeta \nu_{ei}$ with $\zeta=9.5$, and an adjustment in antenna dimension of $R_{\text{sim}}=\xi R_{\text{exp}}$ with $\xi=0.88$. Calculation of the axial gradient of the computed phase variation shows a travelling wave, with a good agreement with data. 

A number of physics reasons, detailed in Section~\ref{subsec: turbulence}, exist to support an enhancement in the collisionality. However, experimental uncertainties in the current, antenna and vessel dimensions are not sufficient to explain the $12\%$ reduction in antenna radius required to find a qualitative match in field amplitude.  A possible omission of the RF solver, that might explain the need for an artificially reduced antenna dimension is the single $m$ vacuum solution limitation. Specifically, EMS can only compute the glass response to the same mode number of the antenna. In reality, however, the antenna will generate $m \neq 1$ vacuum harmonics.  It is possible that the evanescent length scales of $m=-1$ and $m=0$ are much larger than $m=1$, producing substantial $m=-1$ and $m=0$ waves at the plasma-glass surface. Mode coupling at the plasma-glass surface may then couple $m \neq 1$ to $m=1$, producing an increased field in the plasma. 

The local minimum observed both experimentally and numerically in the axial profile of $|B_r|_{\text{rms}}$ around $z=0.27$~m has been also observed in many other devices,\cite{Guo:1999aa, Degeling:2004aa, Light:1995aa, Mori:2004aa} for both uniform and non-uniform field cases. For the uniform field case, it has been suggested that the spatial modulation of the helicon wave amplitude is not caused by reflections from the end boundaries, but by a simultaneous excitation of two radial modes.\cite{Chen:1996ab, Chen:1996aa, Light:1995aa, Mori:2004aa} Similarly, the minimum observed in MAGPIE cannot be explained by standing waves, because the amplitude becomes much smaller at bigger $z$ (suggesting strong damping), and the phase advances with increasing $z$ (denoting a travelling wave). Further, radial profiles of the wave field in Fig. 5 feature a possible superposition of the first and second radial modes of the $m=1$ azimuthal mode. Therefore, we speculate that the minimum observed here may be also due to the simultaneous excitation of two fundamental radial modes. We will show later that radial gradient in plasma density is essential for the excitation of this local minimum under the present experimental conditions.  

\begin{figure*}[ht]
\begin{center}$
\begin{array}{lll}
(a)&\hspace{-0.1cm}(c)&\hspace{-0.1cm}(e)\\
\hspace{-0.05cm}\includegraphics[width=0.33\textwidth,angle=0]{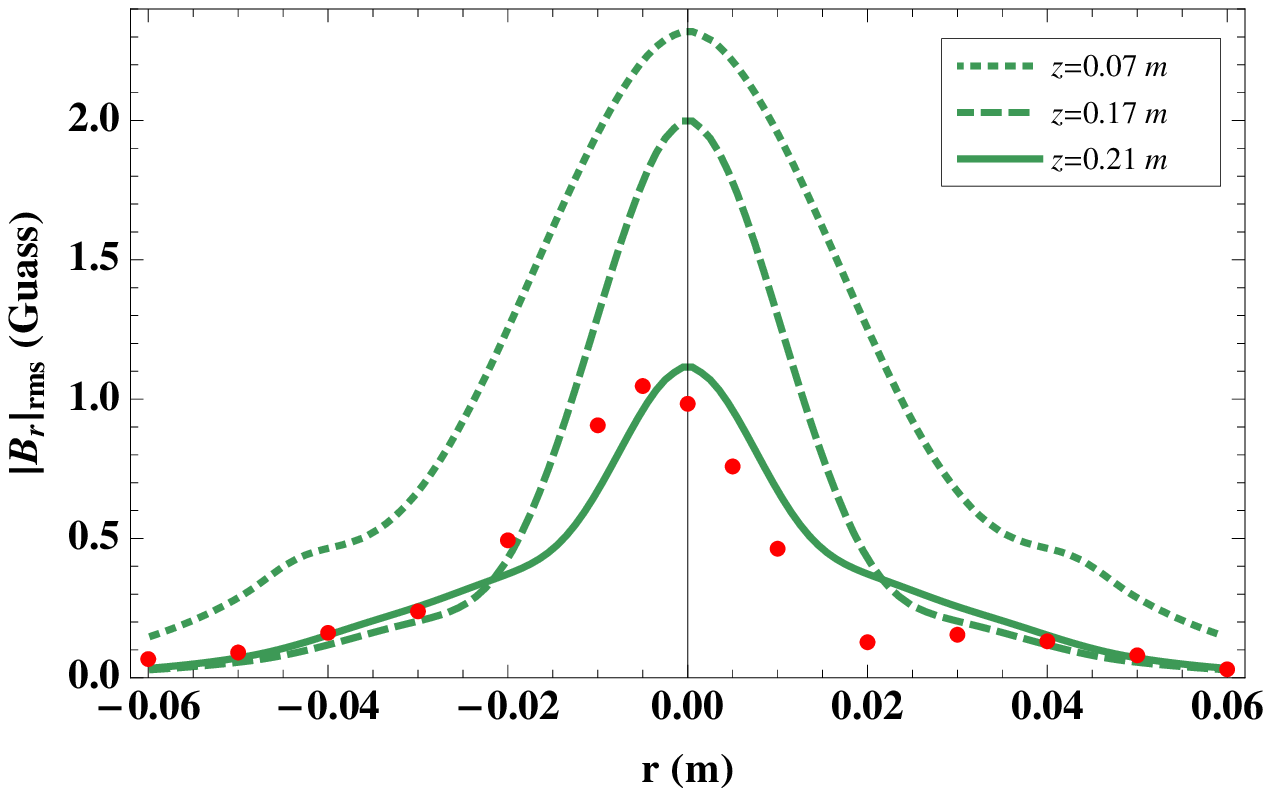} & \hspace{-0.1cm}\includegraphics[width=0.33\textwidth,angle=0]{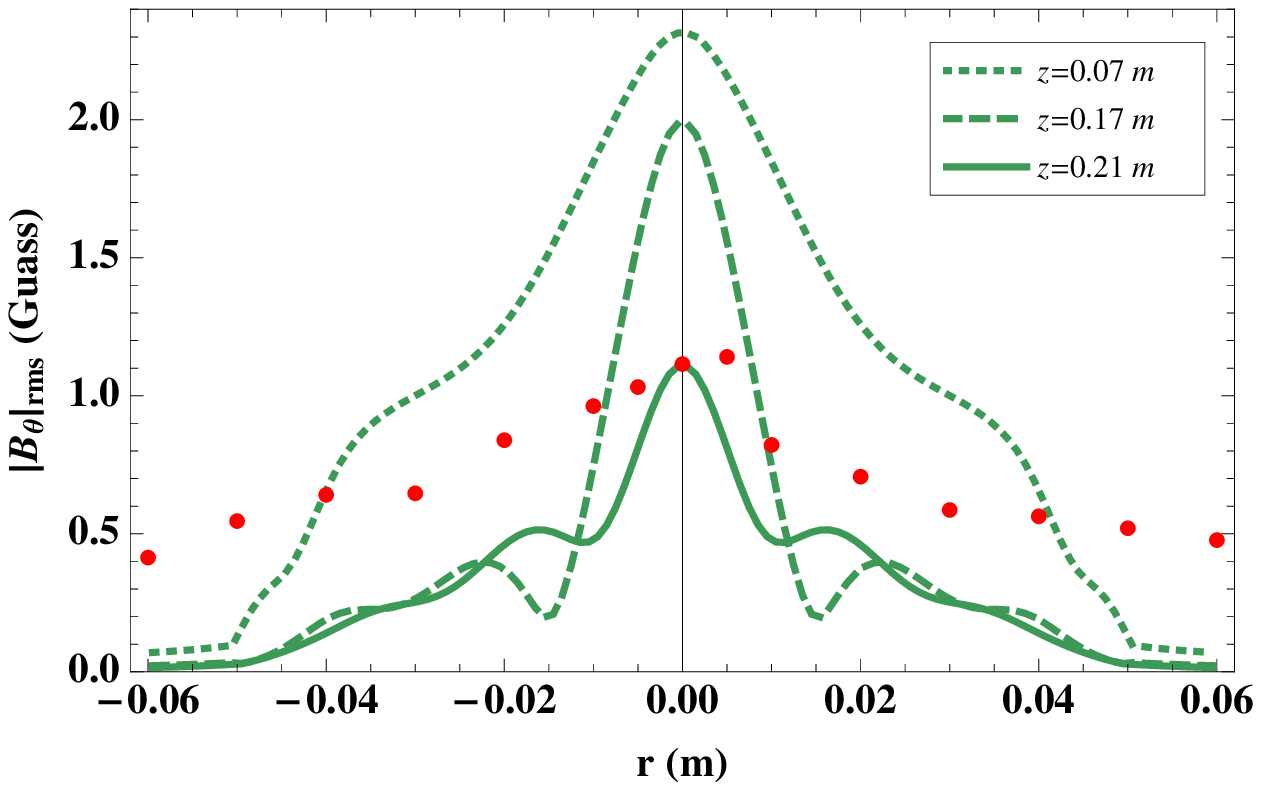} & \hspace{-0.1cm}\includegraphics[width=0.33\textwidth,angle=0]{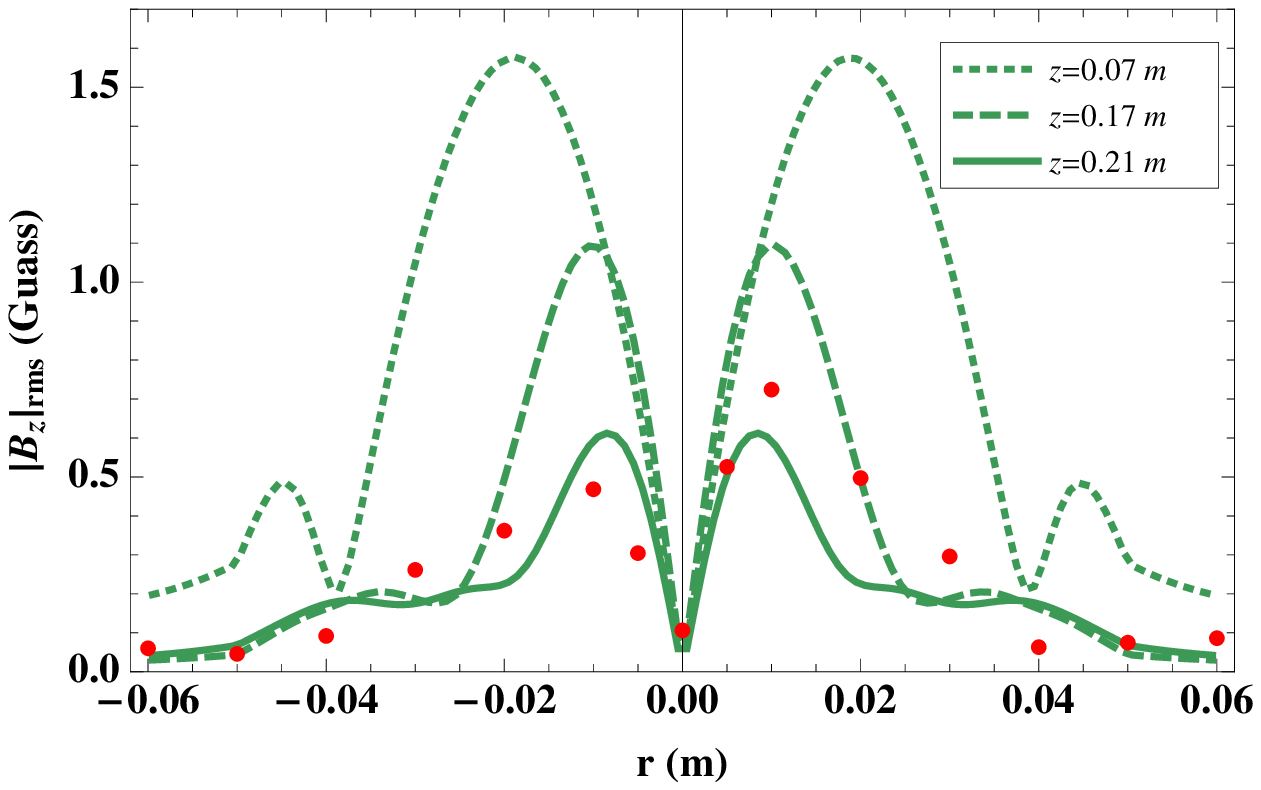} \\
(b)&\hspace{-0.1cm}(d)&\hspace{-0.1cm}(f)\\
\hspace{-0.2cm}\includegraphics[width=0.33\textwidth,angle=0]{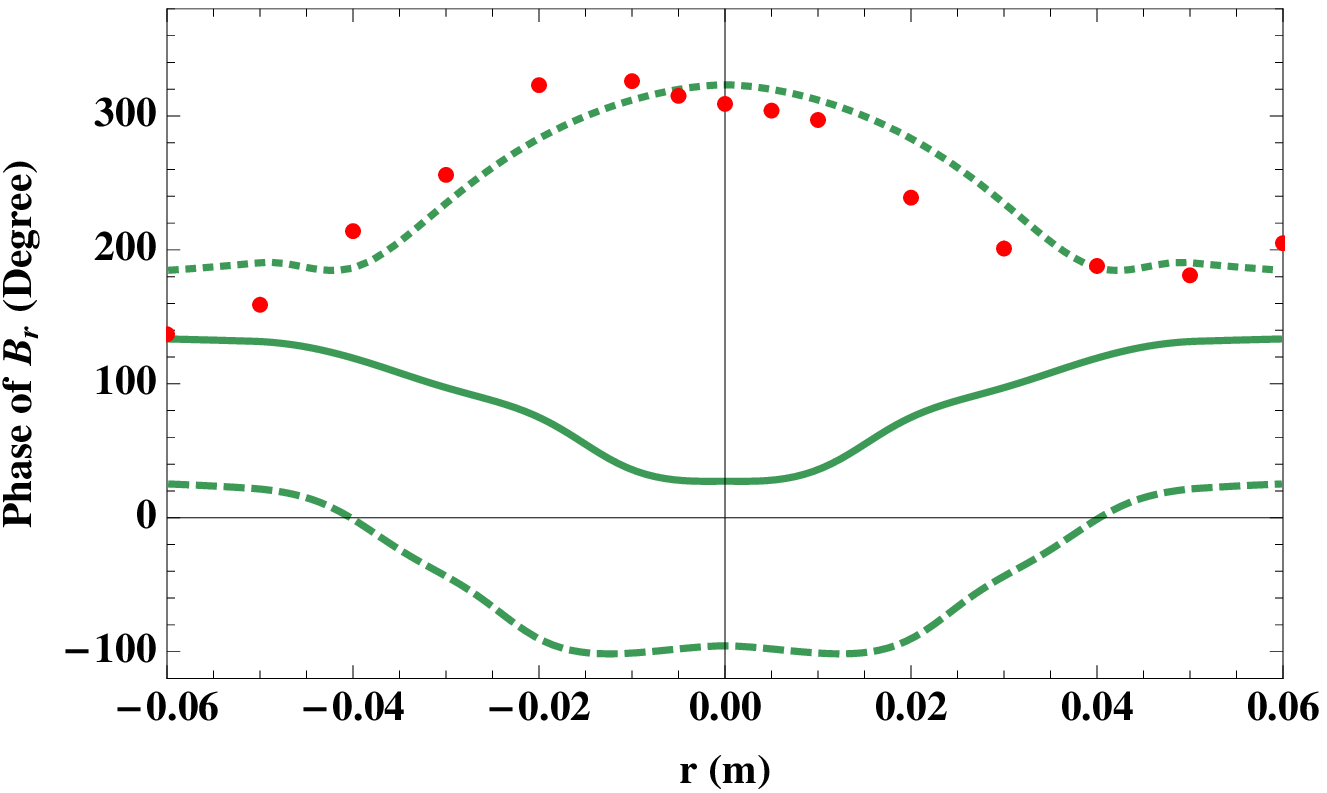} & \hspace{-0.12cm}\includegraphics[width=0.32\textwidth,angle=0]{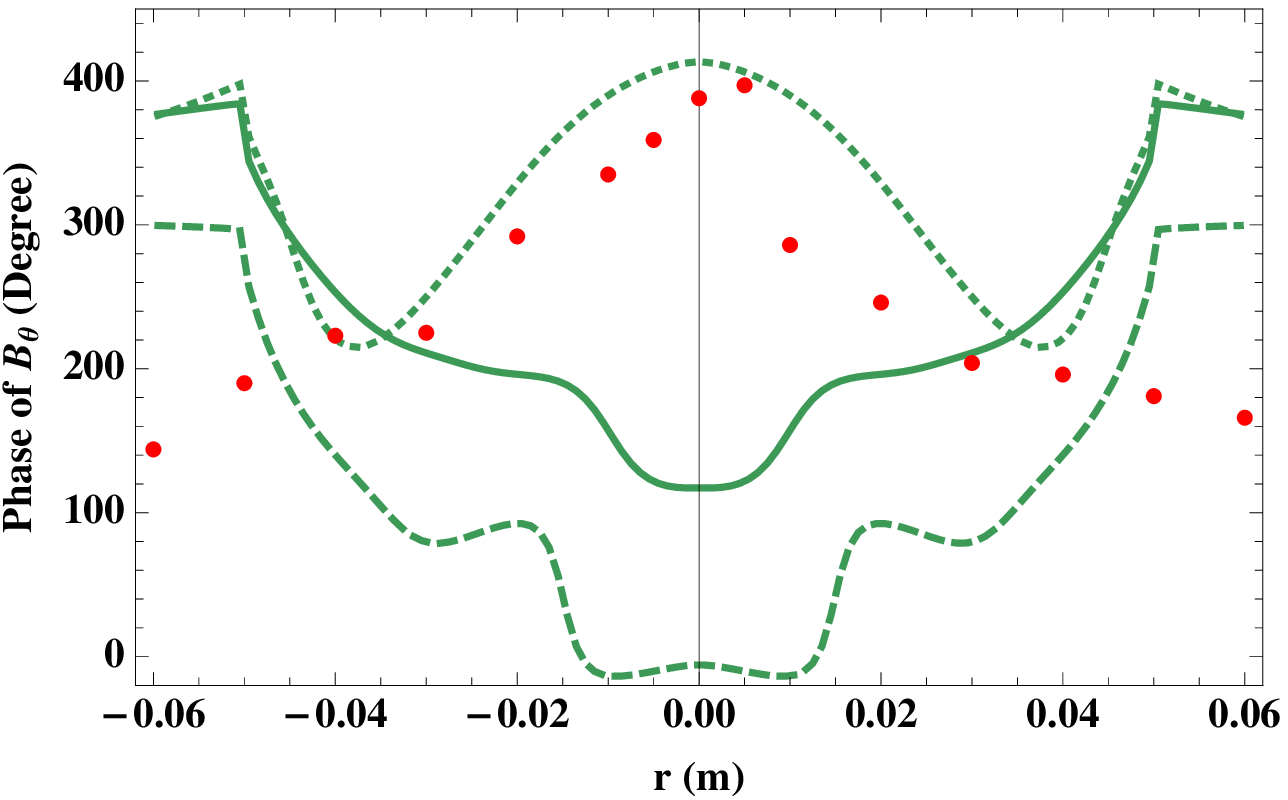} & \hspace{-0.3cm}\includegraphics[width=0.33\textwidth,angle=0]{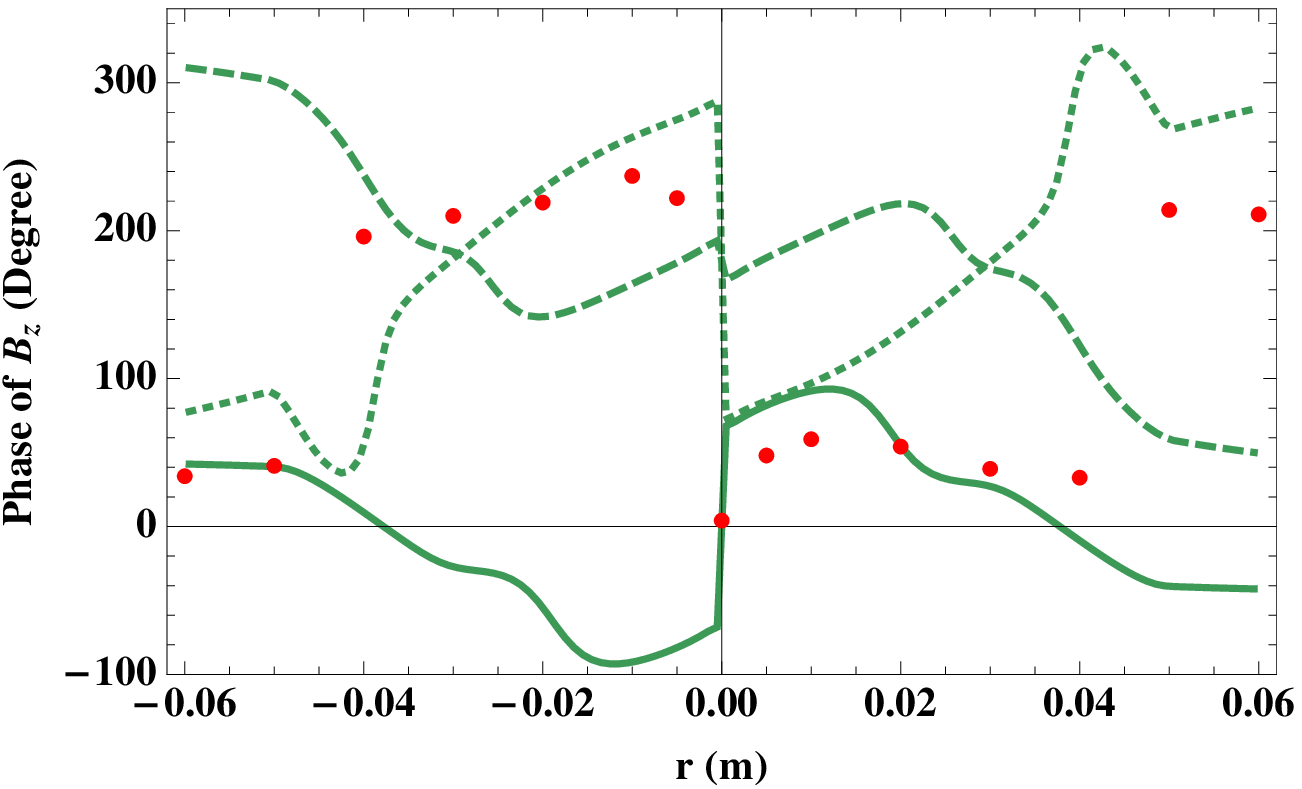} \\
\end{array}$
\end{center}
\caption{Variations of magnetic wave field in radial direction: (a), (c) and (e) are $|B_r|_{\text{rms}}$, $|B_\theta|_{\text{rms}}$ and $|B_z|_{\text{rms}}$, respectively; (b), (d) and (f) are the corresponding phase variations. Dots are experimental data while lines (dotted: $z=0.07$~m, dashed: $z=0.17$~m, solid: $z=0.21$~m) are simulated results. }
\label{radial}
\end{figure*}

Figure 5 shows the radial profiles of computed wave fields for $\nu_{\text{eff}}=9.5 \nu_{ei}$ and $R_{\text{sim}}=0.88R_{\text{exp}}$ at three axial positions in the target region, together with the experimental data measured at $z=0.17$~m. The predicted wave field amplitude profile at $z=0.17$~m is consistent with the data, but the magnitude is nearly double the measured value, and the phase profile is a poor match. We have also computed the wave fields at axial locations with best agreement to the amplitude ($z=0.21$~m) and phase ($z=0.07$~m). We justify this freedom of choice by the experimental uncertainty in axial density profile, and the numerical sensitivity identified in the radial profile of the wave field with axial position. Inspection of Fig. 5 reveals that it is possible to find a reasonable agreement to the wave amplitude and phase profile-albeit-independently. As expected, all calculations show the wave mode structure of $m=1$ through $|B_z(r=0)|_{\text{rms}}\approx 0$, consistent with the antenna parity. The near-null minima in both $|B_\theta(r)|_{\text{rms}}$ and $|B_z(r)|_{\text{rms}}$ suggest a likely simultaneous excitation of the first and second radial modes. This may account for the minimum in Fig. 4(a), and is consistent with conclusions from others.\cite{Chen:1996ab, Mori:2004aa, Chen:1996aa, Light:1995aa, Light:1995ab} Mori et al.\cite{Mori:2004aa} suggested that the superposition feature is associated with wave focusing effect caused by the non-uniform magnetic field in the target region. 

\section{Numerical profile scans}
It has previously been shown that the plasma density can be further increased by introducing a cusp or non-uniform static magnetic field in the vicinity of the helicon antenna.\cite{Boswell:1997aa, Chen:1992aa, Chen:1997aa, Gilland:1998aa} To shed light on the increased plasma production, we perform a detailed numerical study on the effects of radial and axial plasma density gradients and axial magnetic field gradient on wave propagation characteristics. The enhancement of $\zeta=9.5$ to $\nu_{ei}$ and the adjustment of $\xi=0.88$ to $R_{\text{exp}}$ are still employed in this section because they provide a good agreement with the measured wave field. 

\subsection{Axial profile of plasma density}
\label{subsec: density}

\begin{figure}[ht]
\begin{center}$
\begin{array}{c}
\includegraphics[width=0.5\textwidth,angle=0]{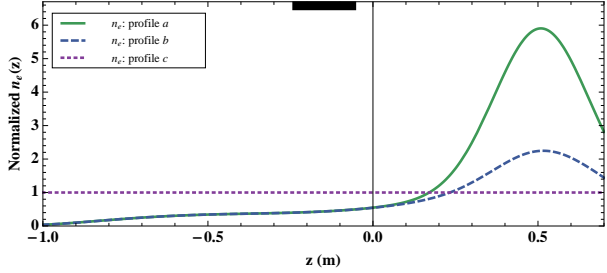}
\end{array}$
\end{center}
\caption{Normalized axial profiles of plasma density on axis. The solid line is associated with density profile linear with $B_0(z)$. The dashed line is the same to the solid one except in region of $0<z<0.7$~m where the density maximum is adjusted to match preliminary experimental observations. The dotted line represents a $z$-independent density profile. }
\label{aden}
\end{figure}

We first study the effect of axial gradient in plasma density, which has been assumed to be linear with the static magnetic field so far, on wave propagations by comparing the wave fields from three different on-axis density profiles shown in Fig. 6. Other conditions are kept the same as previous sections. 

\begin{figure}
\begin{center}$
\begin{array}{l}
(a)\\
\includegraphics[width=0.5\textwidth,angle=0]{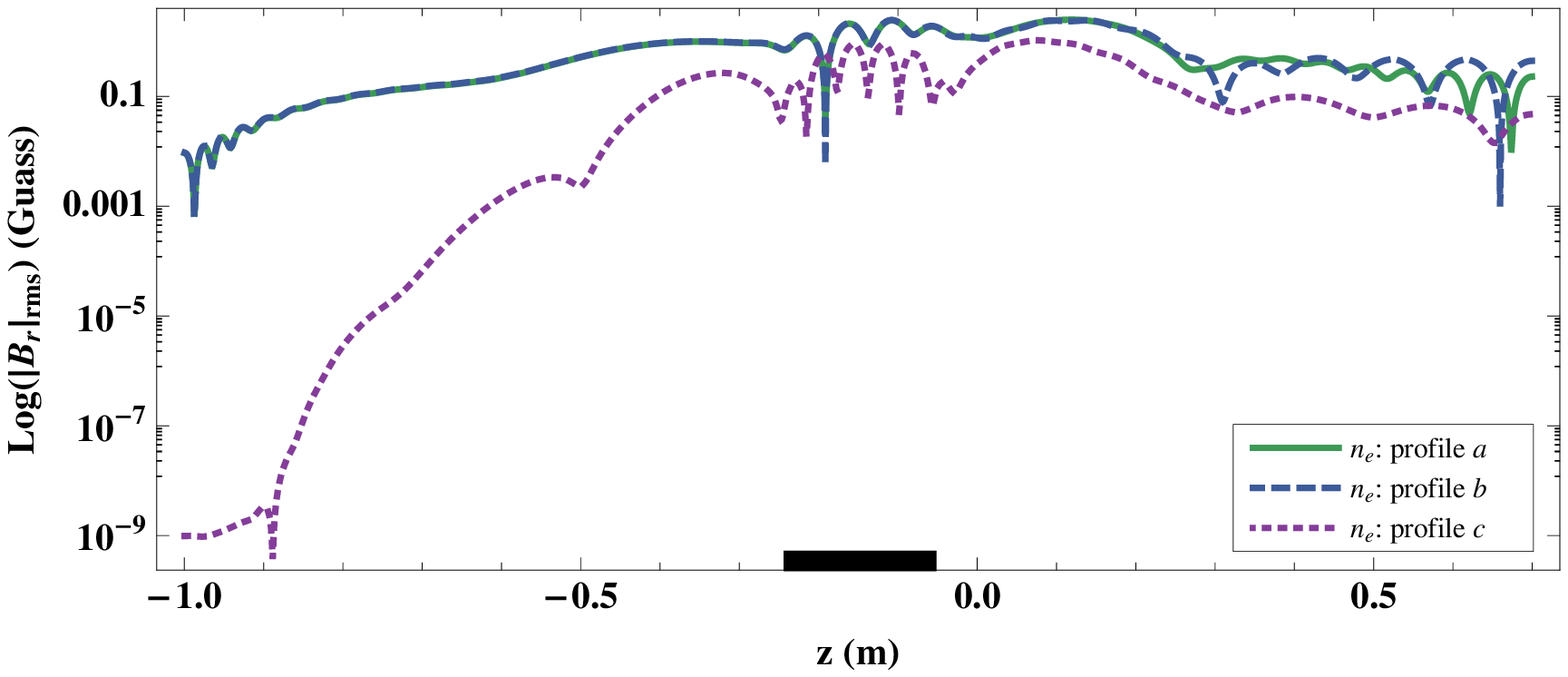}\\
(b)\\
\includegraphics[width=0.468\textwidth,angle=0]{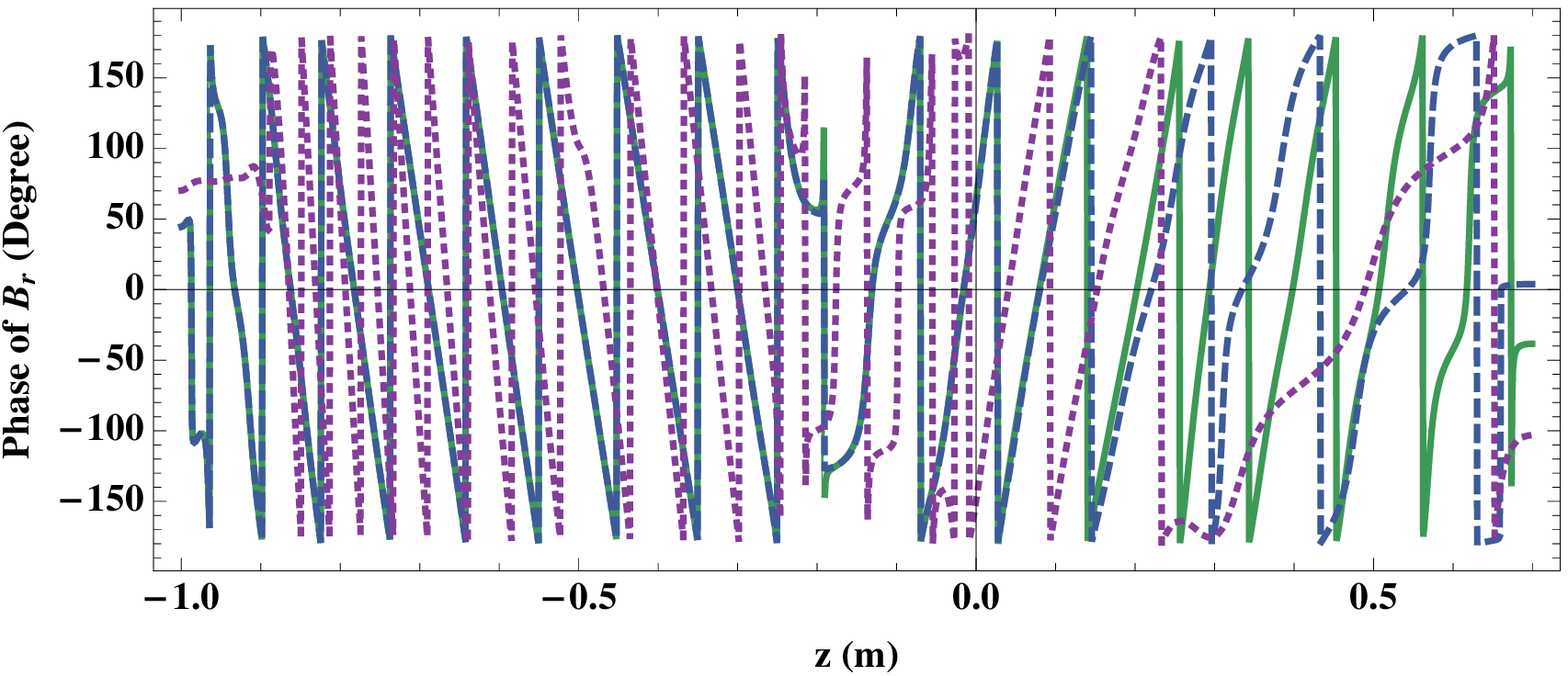}
\end{array}$
\end{center}
\caption{Variations of magnetic wave field in axial direction (on-axis) with $n_e$ axial profile: (a), log scale of $|B_r|_{\text{rms}}$; (b), phase of $B_r$. The three density profiles employed correspond to those in Fig. 6.}
\label{axialden}
\end{figure}

The computed wave fields in axial direction (on-axis) are shown in Fig. 7. A log scale in the amplitude has been employed to see the wave propagation features clearly. We can see from the phase variations (Fig. 7(b)) that as density is decreased in the target region the wavelength increases, which is consistent with a simple theory developed previously, \cite{Chen:1996ab} 
\begin{equation}
\frac{3.83}{R_p}=\frac{\omega}{k}\frac{n_e e \mu_0}{B_0}.
\end{equation}
Thus, if $\omega$, $R_p$ (plasma radius) and $B_0$ are all fixed, $k$ behaves proportional to $n_e$, which means that the wavelength becomes larger at lower density. Here, the value of $3.83$ is the first non-zero Bessel root of $J_1(r)=0$, representing the first radial mode, which is assumed to be dominant in our case. Further, for all density profiles shown in Fig. 6, the wavelength is bigger in region of $0<z<0.6$~m than that in other regions, indicating an increased phase velocity. This increased phase velocity together with the strong decay of wave amplitude suggest strong coupling of RF power from the antenna into the plasma in this region.\cite{Guo:1999aa}

\subsection{Axial profile of static magnetic field}
\label{subsec: access}

\begin{figure}[ht]
\begin{center}$
\begin{array}{c}
\includegraphics[width=0.5\textwidth,angle=0]{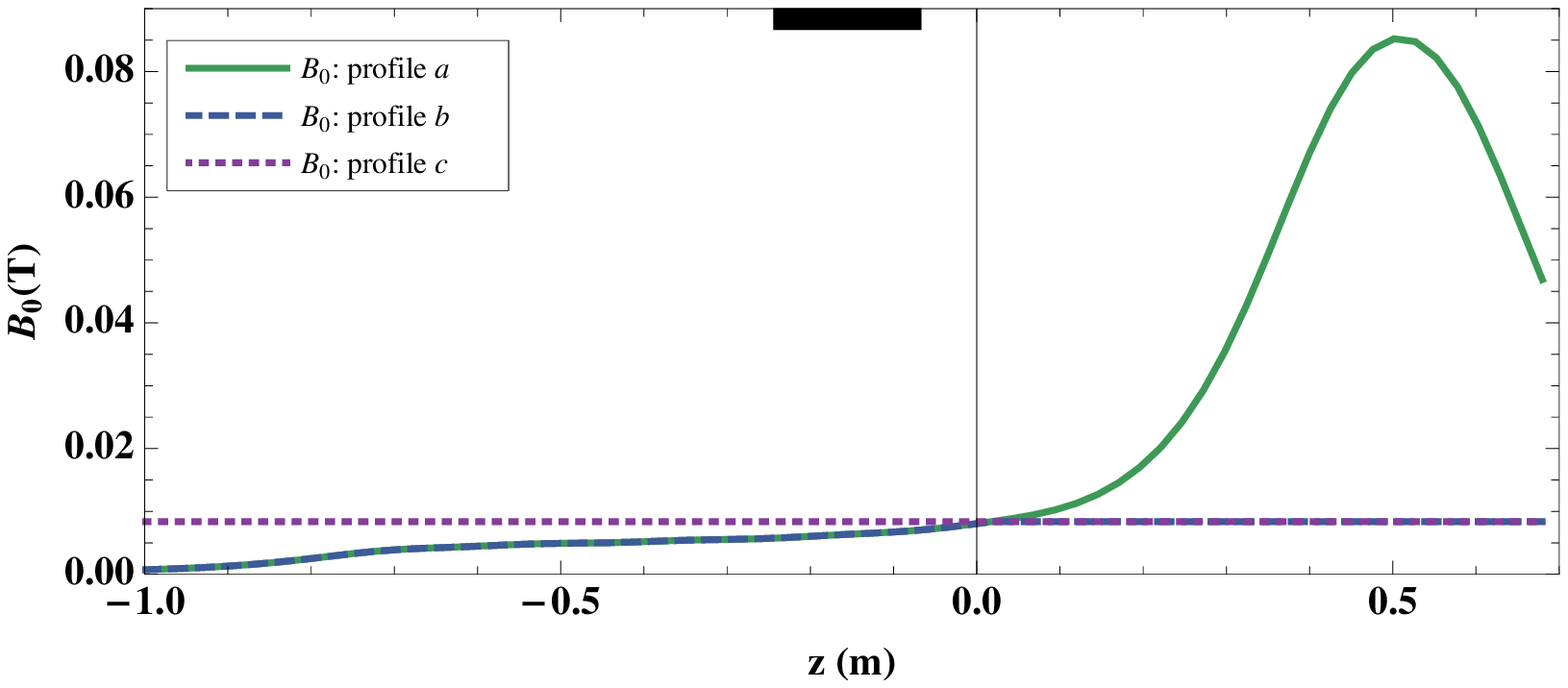}
\end{array}$
\end{center}
\caption{Axial profiles of static magnetic field. The solid line shows original experimental data (Fig. 2(a)), while the dashed line shows the same except being flattened in region of $z>0$~m and the dotted line is flattened everywhere.}
\label{field}
\end{figure}

Second, following section~\ref{subsec: density}, we keep the axially uniform density profile and study the effects of axial gradient in static magnetic field, which is radially near uniform according to Eq. (7). Three employed field profiles are shown in Fig. 8, which enable us to study the effects of field gradient in target and source regions seperately on wave propagations. Comparison between solid and dashed lines in Fig. 9 shows that axial gradient in magnetic field in the target region increases the propagation distance of helicon waves, consistent with Mori et al.'s conclusion that a focused non-uniform magnetic field provides easier access for helicon wave propagations than a uniform field.\cite{Mori:2004aa} The simple theory in Eq. (14) is satisfied again here: with decreased field strength in the target region, the wavelength becomes shorter. Although the difference between dashed and dotted field profiles is small as shown in Fig. 8, the computed field amplitudes are significantly different. Figure 9 shows that with the uniform field profile wave amplitude is much bigger than that with non-uniform field profile for $z<0$~m. Furthermore, waves keep their travelling features till the left endplate for uniform $B_0$, whereas for non-uniform $B_0(z)$ the wavelength becomes smaller when approaching left, and the waves are not travelling at all when $B_0(z)$ is low enough ($z<-0.9$~m). 

\begin{figure}
\begin{center}$
\begin{array}{l}
(a)\\
\includegraphics[width=0.5\textwidth,angle=0]{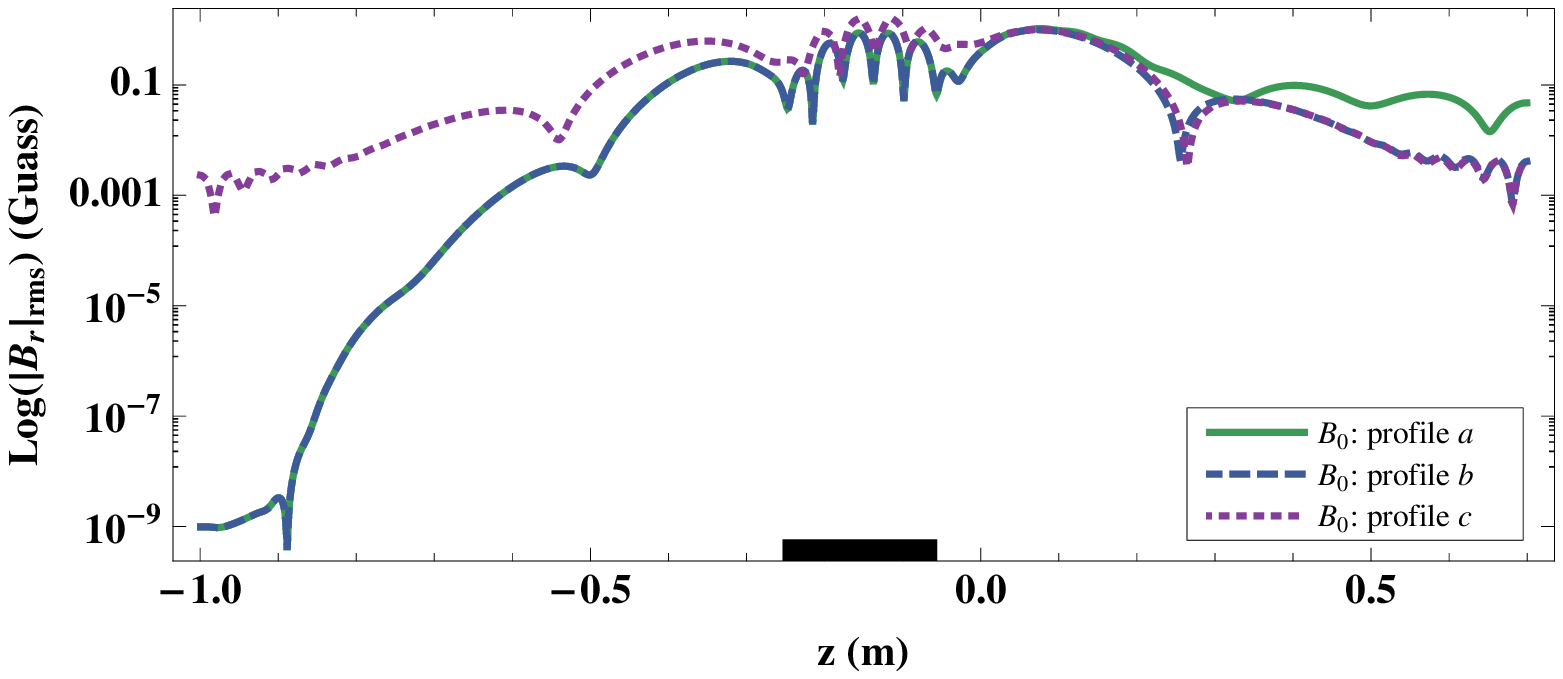}\\
(b)\\
\hspace{0.07 cm}\includegraphics[width=0.46\textwidth,angle=0]{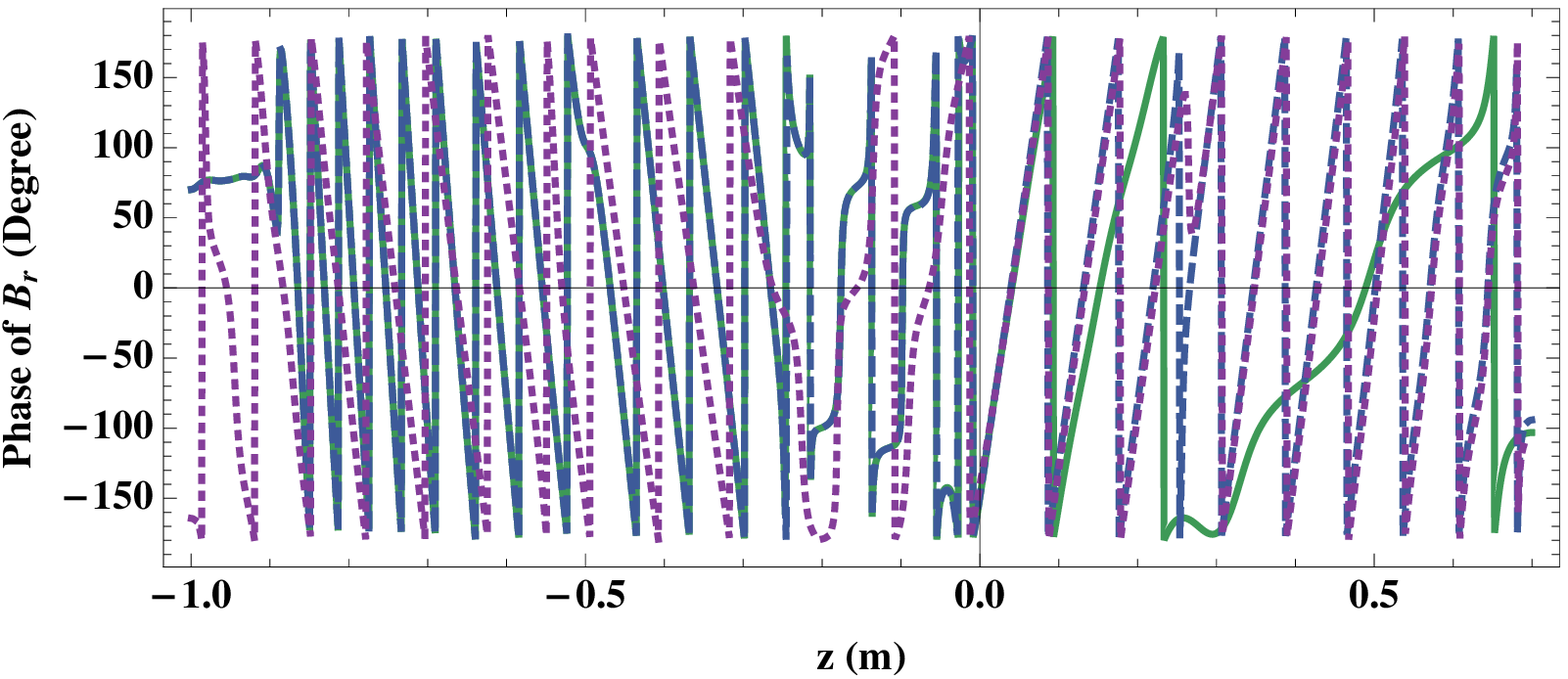}
\end{array}$
\end{center}
\caption{Variations of magnetic wave field in axial direction (on-axis) with $B_0$ axial profile: (a), log scale of $|B_r|_{\text{rms}}$; (b), phase of $B_r$. The three field profiles employed correspond to those in Fig. 8. }
\label{afield}
\end{figure}

\subsection{Radial profile of plasma density}
\label{subsec: rlh}

\begin{figure}
\begin{center}$
\begin{array}{c}
\includegraphics[width=0.4\textwidth,angle=0]{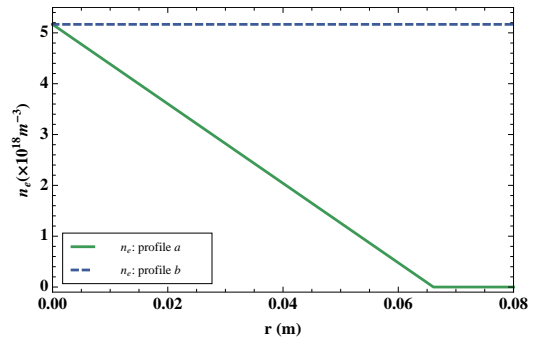}
\end{array}$
\end{center}
\caption{Radial profiles of plasma density. }
\label{rden}
\end{figure}

Now, we keep the plasma density and static magnetic field both uniform in the axial direction, and study the effects of radial gradient in plasma density. The two density profiles employed are shown in Fig. 10, with and without radial gradient, and their corresponding results are shown in Fig. 11. We can see first that the local minimum in wave amplitude profiles, e. g. at $z=-0.52$~m and $z=0.27$~m, disappear when the radial density profile is flat, suggesting that the radial gradient in plasma density is essential to have a local minimum under the present conditions. Second, the wave amplitude is much bigger in both target and source regions for plasma density with radial gradient, suggesting that a radial gradient in density may be useful to maximize the plasma production. 

\begin{figure}
\begin{center}$
\begin{array}{l}
(a)\\
\includegraphics[width=0.5\textwidth,angle=0]{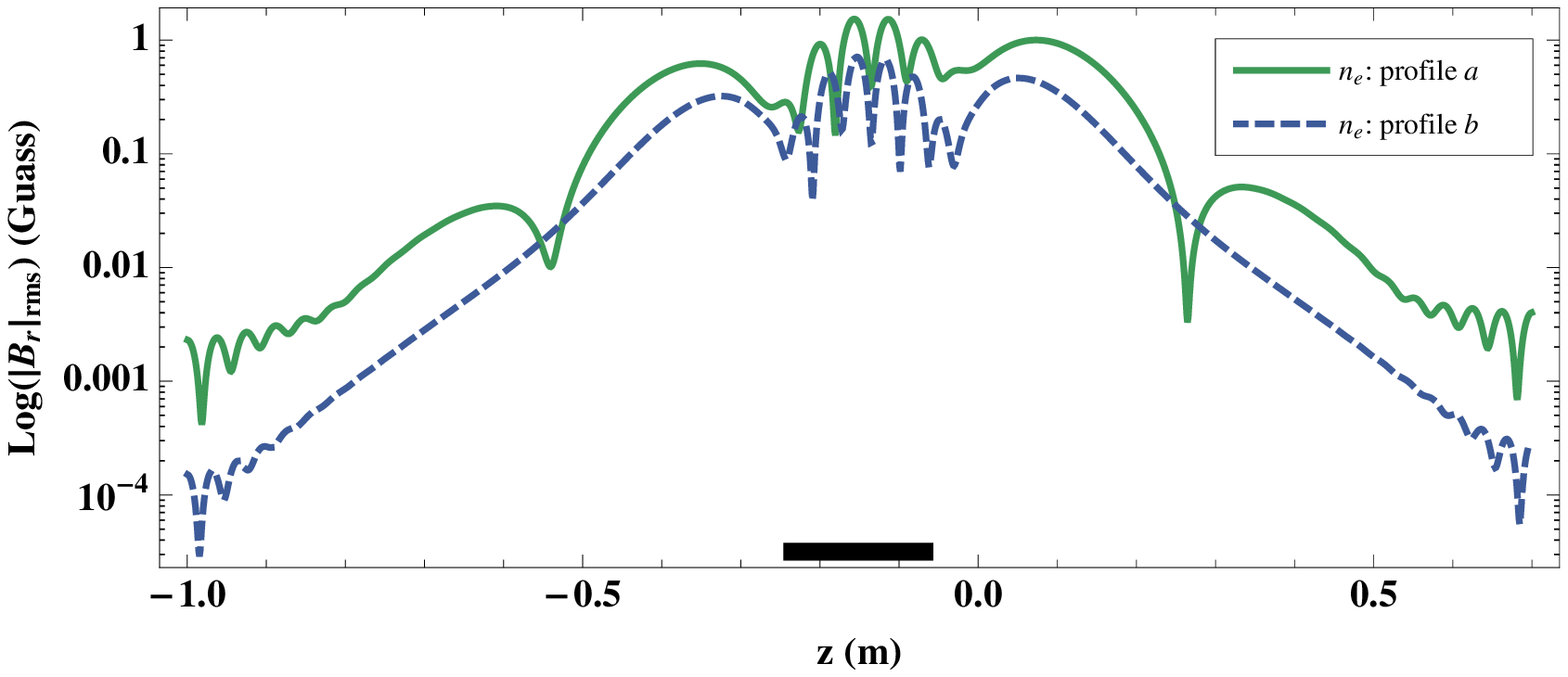}\\
(b)\\
\hspace{0 cm}\includegraphics[width=0.47\textwidth,angle=0]{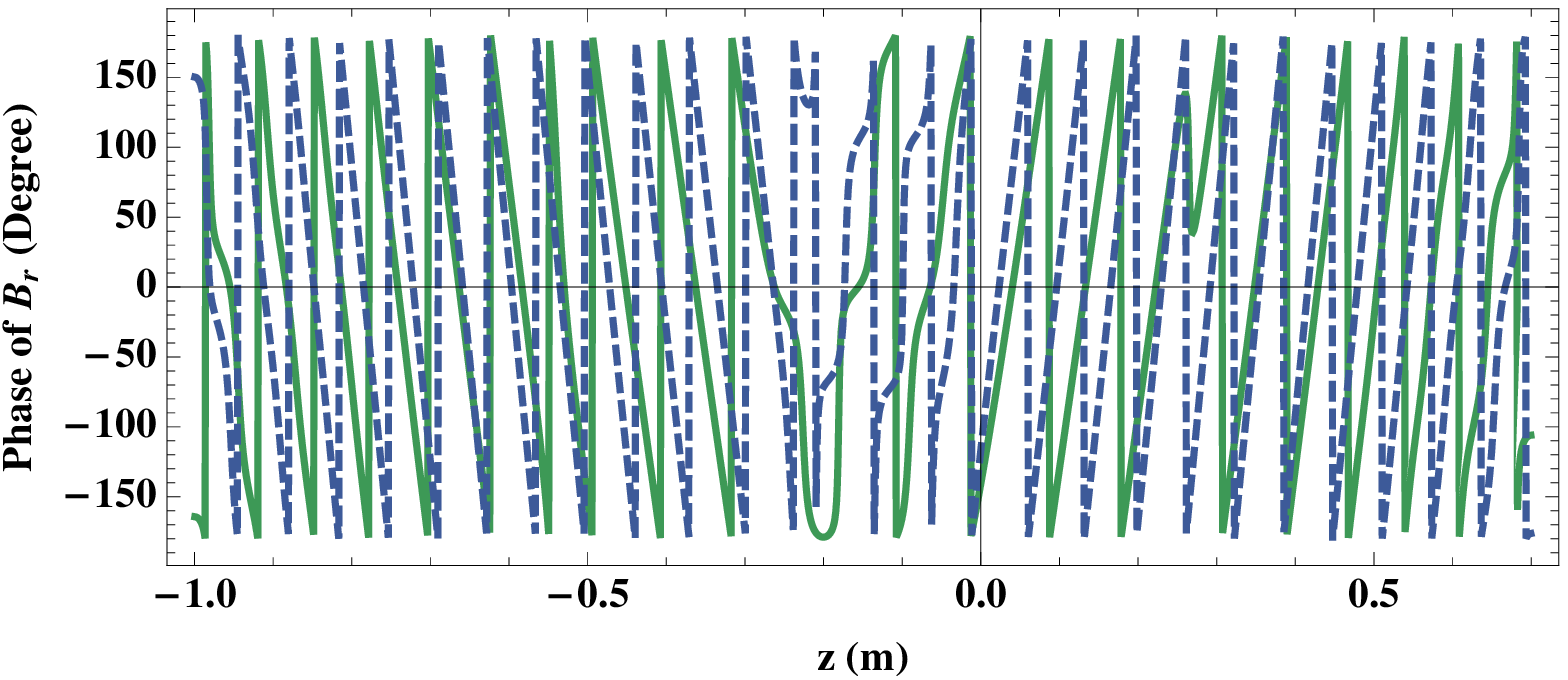}
\end{array}$
\end{center}
\caption{Variations of magnetic wave field in axial direction (on-axis) with $n_e$ radial profile: (a), log scale of $|B_r|_{\text{rms}}$; (b), phase of $B_r$. The two density profiles employed correspond to those in Fig. 10.}
\label{radialden}
\end{figure}

\section{Collisionality and field direction}
\label{subsec: turbulence}

\subsection{Enhancement of electron-ion collision frequency}
In a similar manner to other work,\cite{Zhang:2008aa, Lee:2011aa} we have used an enhancement to $\nu_{ei}$ (here $\nu_{\text{eff}}\approx 9.5\nu_{ei}$), in order to find a qualitative match of simulated wave field to the data. In this section, we explore the physical consequences of scaling $\nu_{\text{eff}}/\nu_{ei}$ in simulations while keep using the adjustment $\xi=0.88$ in the antenna radius. 

\begin{figure}
\begin{center}$
\begin{array}{l}
\includegraphics[width=0.5\textwidth,angle=0]{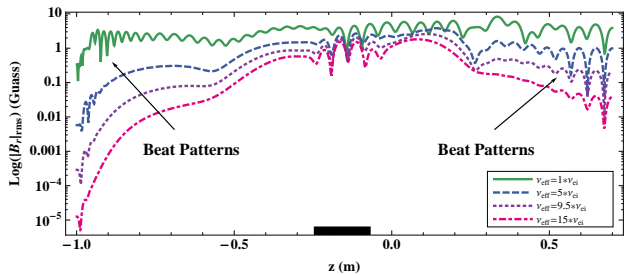}\\
\end{array}$
\end{center}
\caption{Variations of on-axis wave amplitudes in axial direction with different electron-ion collision frequencies: log scale of $|B_r|_{\text{rms}}$.}
\label{collision}
\end{figure}

Variations of wave amplitude on axis in the axial direction for different collision frequencies are shown in Fig. 12. As the collision frequency increased from $\nu_{ei}$ to $15\nu_{ei}$, the wave amplitude decreases nearly everywhere, and the wave decay length is shortened. This indicates that the wave energy or power coupled from the antenna to the core plasma drops as the collision frequency becomes higher, and the power is more absorbed under the antenna. This is consistent with a previous conclusion that the RF energy is almost all absorbed in the near region of the antenna rather than in the far region.\cite{Chen:1996aa} The oscillations near the downstream and upstream ends at low $\nu_{\text{eff}}$ are caused by reflections from the ideally conducting endplates, which disappear if the endplates are moved further away.

As suggested by Lee et al.,\cite{Lee:2011aa} an enhanced electron-ion collision frequency may be due to ion-acoustic turbulence which can happen if the electron drift velocity exceeds the speed of sound in magnetized plasmas. Based on the experimental conditions in MAGPIE, we have calculated the threshold field strength $B_{\text{T}}$, below which ion-acoustic turbulence can happen. This threshold is given by $v_D \geq C_s$, where $v_D \approx k_B T_e/|e| B_0 R_p$ is the electron drift velocity with $k_B$ Boltzmann's constant and $C_s=\sqrt{k_B T_e/m_i} $ the speed of sound in magnetized plasmas, resulting in $B_{\text{T}} \leq 0.0224~\text{T}$. Thus, the whole source region which produces helicon plasmas and waves is indeed located within this range. The ion-acoustic turbulence has the effect of providing additional electron-ion collisions within a dielectric tensor model, and thereby improves the agreement with observations. 

\subsection{Direction of static magnetic field}
Observations have been made previously that the directionality of helicon wave propagations is dependent on the direction of static magnetic field in helicon discharges using helical antennas, but all in uniform field configurations.\cite{Chen:1996aa, Chen:1996ab, Sudit:1996aa, Lee:2011aa} In this section, we study the directionality in a non-uniform field configuration. Specifically, we have computed the wave amplitude and wave energy density in MAGPIE for the experimental and field reversed configurations. In MAGPIE, the field points from target to source, as mentioned in section~\ref{subsec: antenna}. Figure 13 shows the computed axial profiles of wave amplitude on axis and 2D contour plots of wave energy density for both field direction pointing from target to source (Fig. 13(a) and 13(b)) and field direction pointing from source to target (Fig. 13(c) and 13(d)). In this calculation, we have chosen $\nu_{\text{eff}}=\nu_{ei}$ to see more details, and chosen the density profile to be linear with $B_0(z)$ in the axial direction and non-uniform in radius as measured in Fig. 2(b). The field strength profile used here is shown in Fig. 2(a). Inspection of Fig. 13 reveals that the wave energy is larger in the opposite side of the antenna, relative to the direction of the static magnetic field. This observation has been confirmed experimentally through finding that the plasma is brighter in the opposite side of the antenna relative to the direction of the applied external field. Therefore, the dependence of the direction of helicon wave propagations to that of static magnetic field still exists even when the field configuration is non-uniform.

\begin{figure*}
\begin{center}$
\begin{array}{ll}
(a)&(c)\\
\hspace{0.25 cm}\includegraphics[width=0.433\textwidth,angle=0]{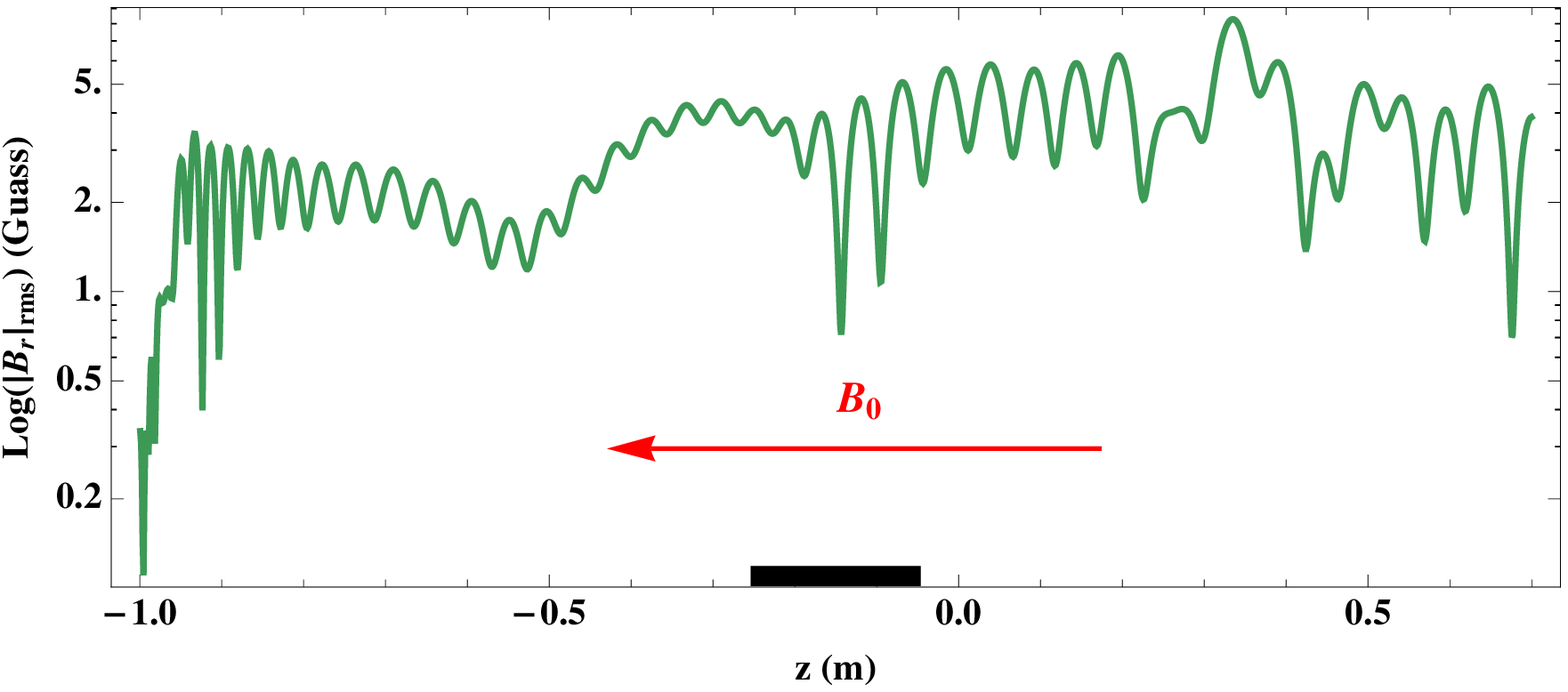}&\hspace{0 cm}\includegraphics[width=0.428\textwidth,angle=0]{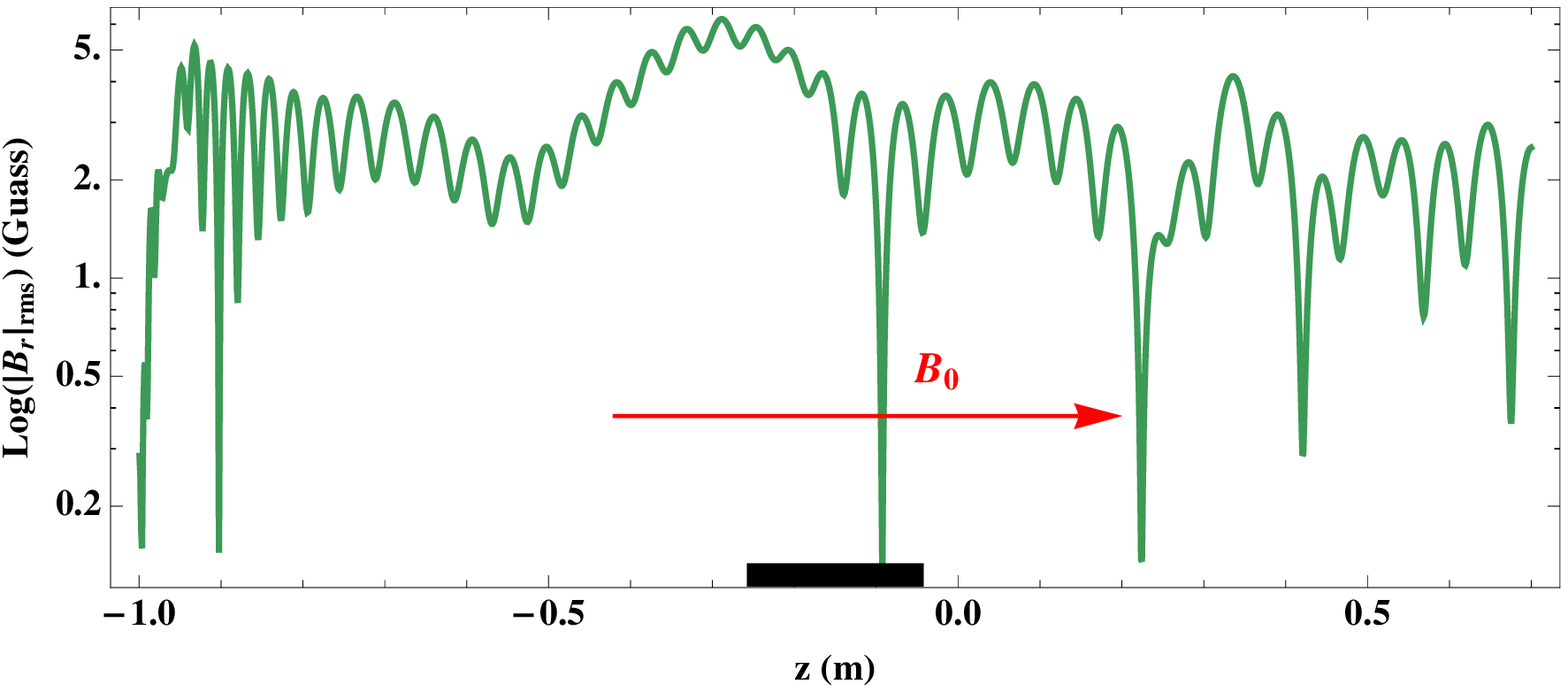}\\
(b)&(d)\\
\hspace{0 cm}\includegraphics[width=0.495\textwidth,angle=0]{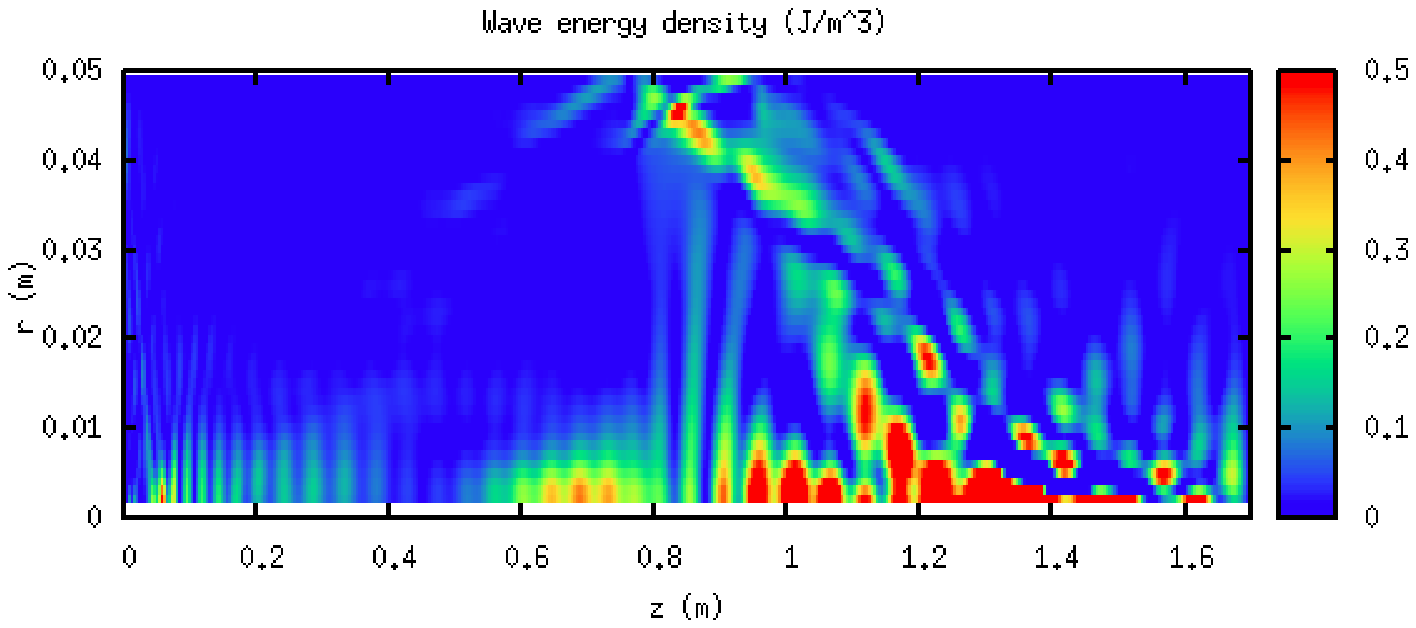}&\hspace{-0.23 cm}\includegraphics[width=0.495\textwidth,angle=0]{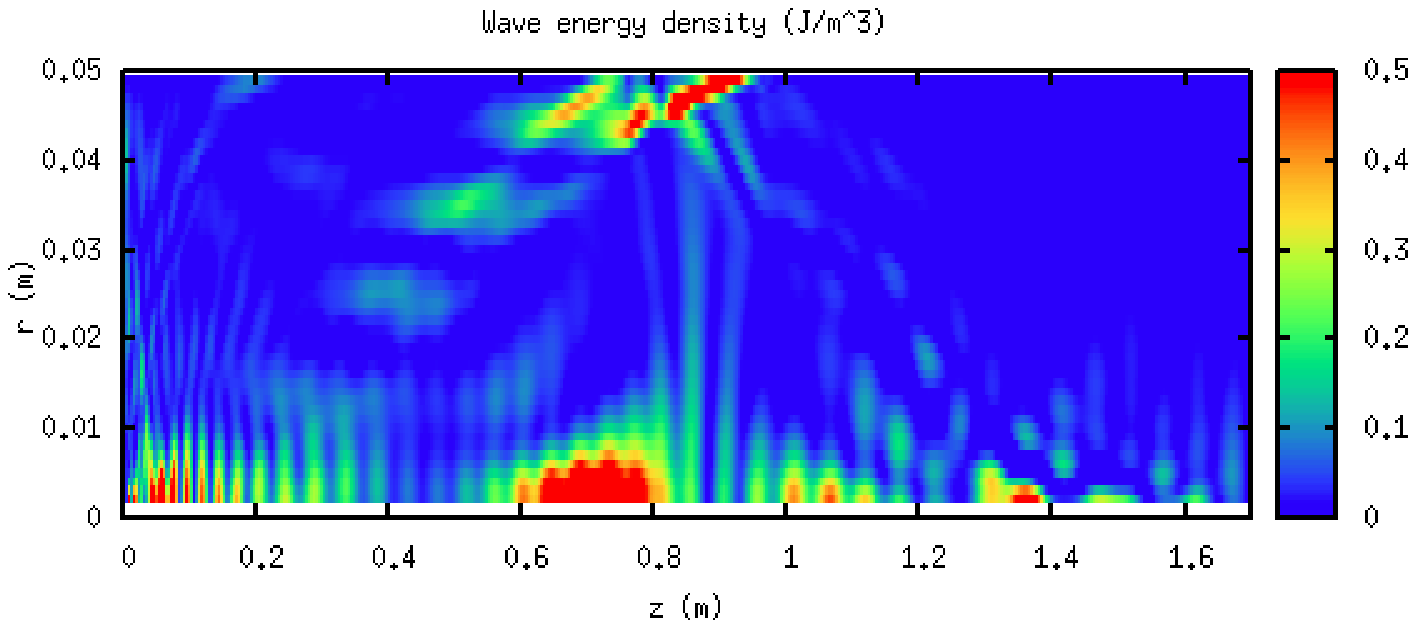}
\end{array}$
\end{center}
\caption{Axial profiles of magnetic wave field (on-axis) and contour plot of wave energy density in (z, r) space for a non-uniform plasma density: (a), log scale of $|B_r|_{\text{rms}}$ for upstream $B_0(z)$; (b), wave energy density for upstream $B_0(z)$; (c), log scale of $|B_r|_{\text{rms}}$ for downstream $B_0(z)$; (d) wave energy density for downstream $B_0(z)$.}
\label{ndirec}
\end{figure*}

\section{conclusions}
A RF field solver based on Maxwell's equations and a cold plasma dielectric tensor is employed to describe the wave phenomena observed in a cylindrical non-uniform helicon discharge, MAGPIE. Here, the non-uniformity is both radial and axial: the plasma density is dependent on $r$ and $z$, the static magnetic field varies with $z$, and the electron temperature is a function of $r$. A linear fitting was conducted for radial profiles of plasma density and electron temperature, and the fitted profiles utilized in wave field calculations. A linear relationship between the axial profile of plasma density and the static magnetic field was assumed. Other conditions used in the simulation were taken from experiment directly, including filling gas (argon), antenna current $38.8$~A (magnitude), driving frequency $13.56$~MHz, and a left hand half-turn helical antenna. 

With an enhancement factor of $9.5$ to the electron-ion Coulomb collision frequency $\nu_{ei}$ and $12\%$ reduction in the antenna radius, the wave solver produced consistent wave fields compared to experimental data, including the axial and radial profiles of wave amplitude and phase. Particularly, a local minimum in the axial profiles of wave amplitude was observed both experimentally and numerically, agreeing with previous studies.\cite{Guo:1999aa, Degeling:2004aa, Light:1995aa, Mori:2004aa} Mode structure of $m=1$ is consistent with the left hand half-turn helical antenna being used. A possible explanation for the enhanced electron-ion collision frequency has been offered through ion-acoustic turbulence, which can happen if electron drift velocity exceeds the speed of sound in magnetized plasmas.\cite{Lee:2011aa} Through calculating these two speeds based on MAGPIE conditions, we found that it is indeed satisfied in the source region of MAGPIE where the helicon plasmas and waves are produced. Furthermore, to overcome the single $m$ vacuum solution limitations of the RF solver, which can only compute the glass response to the same mode number of the antenna, we have adjusted the antenna radius to match the wave field strength in the plasma. 

A numerical study on the effects of axial gradients in plasma density and static magnetic field on wave propagations was carried out. This showed that the axial gradient in magnetic field increases the decay length of helicon waves in the target region. The strong decay of wave amplitude and the increase in phase velocity in region of $0<z<0.6$~m indicate strong coupling of RF power from the antenna to the plasma, which is consistent with a previous study by Guo et al..\cite{Guo:1999aa} The relationship between plasma density, static magnetic field and axial wavelength is consistent with a simple theory developed previously.\cite{Chen:1996ab} 

A numerical scan of the enhancement factor to $\nu_{ei}$ reveals that with increased electron-ion collision frequency the wave amplitude is lowered and more focused near the antenna. This is mainly because of stronger edge heating at higher collision frequencies which prevent more energy transported from the antenna into the core plasma. The amplitude profile at $\nu_{\text{eff}}=9.5\nu_{ei}$, which agrees with experimental data, shows consistent feature with a previous study that the RF energy is almost all absorbed in the near region of the antenna rather than in the far region.\cite{Chen:1996aa} We further studied the effect of the direction of static magnetic field on wave propagations, and found the antiparallel feature that waves propagate in the opposite direction of magnetic field. This dependence of the direction of helicon wave propagations to that of static magnetic field in a non-uniform field configuration is consistent with previous observations made in uniform field configurations.\cite{Chen:1996aa, Chen:1996ab, Sudit:1996aa, Lee:2011aa} 

Physics questions raised by this work include: further explanation of exactly how axially non-uniform field might affect the radially localized helicon mode,\cite{Breizman:2000aa} inclusion of different $m$ numbers in the glass layer and any subsequent coupling to the plasma at the plasma-glass interface, and identification of independent first and second radial modes that superpose to yield a local minimum in wave field amplitude at $z=0.27$~m. Experimental measurements that might corroborate the wave field generation mechanism and associated physics include: the measurement of axial profile of density, measurements of $n_e$, $\mathbf{B}$, and $\mathbf{E}$ with a reversed field. 

\begin{acknowledgements}
We would like to thank Prof Boris N. Breizman and Dr Alexey V. Arefiev (Institute for Fusion Studies, The University of Texas at Austin), Dr Michael Fitzgerald and Dr Trevor A. Lafleur (Plasma Research Laboratory, The Australian National University) for many fruitful discussions. One of the authors-Lei Chang-appreciates the financial support provided by Chinese Scholarship Council for his PhD study at The Australian National University, and the Student Conference Support from Australian Institute of Physics for him to present this work in the coming 39th European Physical Society Conference on Plasma Physics and 16th International Congress on Plasma Physics (Stockholm, Sweden, 2-6 July 2012).
\end{acknowledgements}

\bibliographystyle{aipnum4-1}
\bibliography{pop}

\end{document}